\documentclass[structabstract]{aa}

\usepackage{txfonts}
\usepackage{graphicx}
\usepackage{natbib}
\usepackage{nicefrac}
\usepackage{pgfplots}
\usepackage{booktabs}
\usepackage{multirow}
\usepackage{ulem}
\bibpunct{(}{)}{;}{a}{}{,}

\begin{document}

\title{White dwarfs in the building blocks of the Galactic spheroid} 

\author{Pim van Oirschot\inst{1}\thanks{\email{P.vanOirschot@astro.ru.nl}}
  \and Gijs Nelemans\inst{1,2} \and Else Starkenburg\inst{3} \and \\
  Silvia Toonen\inst{4} \and Amina Helmi\inst{5} \and Simon Portegies Zwart\inst{4}} 

\institute{Department of Astrophysics/IMAPP, Radboud University,
  P.O. Box 9010, 6500 GL Nijmegen, The Netherlands 
  \and Institute for Astronomy, KU Leuven, 
  Celestijnenlaan 200D, 3001 Leuven, Belgium
  \and Leibniz-Institut f\"ur Astrophysik Potsdam, AIP, 
  An der Sternwarte 16, D-14482 Potsdam, Germany
  \and Leiden Observatory, Leiden University, 
  P.O. Box 9513, 2300 RA Leiden, The Netherlands
  \and Kapteyn Astronomical Institute, University of Groningen, 
  P.O. Box 800, 9700 AV, Groningen, The Netherlands} 

\date{Received 9 Sep 2016 / Accepted 10 Sep 2016}

\abstract {} {The galactic halo likely grew over time in part by assembling smaller galaxies, the so-called building blocks. We investigate if 
the properties of these building blocks are reflected in the halo white dwarf (WD) population in the Solar neighborhood.
Furthermore, we compute the halo WD luminosity functions (WDLFs) for four major building blocks of
five cosmologically motivated stellar haloes. We compare the sum of these to the
observed WDLF of the galactic halo, derived from selected halo WDs in the SuperCOSMOS Sky Survey, 
aiming to investigate if they match better than the WDLFs predicted by simpler models. } 
{ We couple the SeBa binary population synthesis model to the Munich-Groningen semi-analytic galaxy formation model,
applied to the high-resolution Aquarius dark matter simulations. Although the semi-analytic model assumes an instantaneous
recycling approximation, we model the evolution of zero-age main sequence stars to WDs, taking age and metallicity variations 
of the population into account. To be consistent with the observed stellar halo mass density in the Solar neighborhood ($\rho_0$), 
we simulate the mass in WDs corresponding to this density, assuming a Chabrier initial mass function (IMF) and a binary fraction of 50\%. 
We also normalize our WDLFs to $\rho_0$.} {Although the majority of halo stars is old and metal-poor and therefore the 
WDs in the different building blocks have similar properties (including present-day luminosity), we find in our models that 
the WDs originating from building blocks that have young and/or metal-rich stars can be distinguished from WDs that were born 
in other building blocks. In practice however, it will be hard to prove that these WDs really originate from different 
building blocks, as the variations in the halo WD population due to binary WD mergers result in similar effects.
The five joined stellar halo WD populations that we modelled result in WDLFs that are very similar to each other.
We find that simple models with a Kroupa or Salpeter IMF fit the observed luminosity function slightly better,
since the Chabrier IMF is more top-heavy, although this result is dependent on our choice of $\rho_0$.} {}

\titlerunning{White dwarfs in the building blocks of the Galactic halo}
\authorrunning{P. van Oirschot et al.}
\maketitle 

\section{Introduction}

When aiming to understand the formation and evolution of our Galaxy, its oldest and most metal-poor component,
the Galactic halo, is an excellent place to study. The oldest
stars in our Galaxy are thought to be formed within 200 million years after the Big Bang, 
at redshifts of $\sim 20-30$ \citep{Couchman:1986}. Being formed in the largest over-densities
that grew gravitationally with time, these stars are now expected to be found predominantly in the innermost
regions of the Galactic spheroid, the Galactic bulge \citep{Tumlinson:2010,Salvadori:2010,Howes:2015,Starkenburg:2016}, 
although also a significant fraction will remain in the halo.
It is still unclear whether the most metal poor stars located in the bulge are actually 
part of the thick disc or halo, or that they are part of a distinct `old spheroid' bulge population
\citep{Ness:2013,Gonzalez:2015,Ness:2016}.
Therefore, although the stellar halo and bulge are classically considered to be two distinct components 
of our Galaxy, it is very practical to study them collectively as the stellar spheroid.

In a recent study on the accretion history of the stellar spheroid of the Milky Way 
\citep{van-Oirschot:2017}, we modelled how this composite component grew over time, by assembling smaller galaxies, 
its so called building blocks. Post-processing the cosmological N-body 
simulations of six Milky Way sized dark matter haloes \citep[the Aquarius project;][]{Springel:2008} with 
a semi-analytic model for galaxy formation \citep{Starkenburg:2013}, we investigated building block properties 
such as mass, age and metallicity. In this work, we apply our findings on the build-up of 
the stellar spheroid to a detailed population study of the halo white dwarfs (WDs). In particular, we will investigate 
if there are still signatures of the spheroid's building blocks reflected in today's halo WD population
that can be observed with the Gaia satellite.

In \citet[][hereafter Paper~I]{van-Oirschot:2014} we already modelled a halo WD population assuming a simple star 
formation history of the stellar halo and a single metallicity value ($Z = 0.001$) for all zero age main sequence (ZAMS) 
stars in the halo. Using the outputs of our semi-analytic galaxy formation model, we can now use a more detailed and 
cosmologically motivated star formation history and metallicity values as input parameters for a population study on halo white
dwarfs. Apart from investigating if this more carefully modelled WD population has properties reflecting WD origins in 
different Galactic building blocks, we will compute its halo white dwarf luminosity function (WDLF). The WDLF
is known to be a powerful tool for studying the Galactic halo, since the pioneering works of 
\citet{Adams:1996,Chabrier:1996,Chabrier:1999} and \citet{Isern:1998}. Particularly, the falloff of the number 
of observed WDs below a certain luminosity can be used to determine the age of the population. 

The setup of the paper is as follows: in section~\ref{sec:2} we summarize how we model the accreted spheroid 
of the Milky Way and what its building blocks' properties are. In this section, we will also explain how we disentangle 
building blocks stars that we expect to find in the stellar halo from those that we 
expect to contribute mainly to the innermost regions of the spheroid (i.e. contribute to the Galactic Bulge).
In section~\ref{sec:3} we explain how we model binary evolution, WD cooling and extinction. 
In section~\ref{sec:4} we show how
observable differences in halo WDs occur due to their origins in the various building blocks that 
contribute to the stellar halo in the Solar neighborhood. We investigate the 
halo WDLF of five simulated stellar halo WD populations in section~\ref{sec:5}.
There, we will also discuss how our findings relate to the recent work of \citet{Cojocaru:2015}.
We conclude in section~\ref{sec:6}.

\section{Stellar haloes and their building blocks}~\label{sec:2}

In this paper we focus on the accreted component of the Galactic spheroid. 
We do not consider spheroid stars to be formed in situ, since we assume that 
this only happens during major mergers\footnote{Here, a merger is classified as `major' if the mass ratio
(mass in stars and cold gas) of the merging galaxies is larger than 0.3.}, but none of our modelled Milky Way galaxies 
experienced a major merger. 
Stellar spheroids also grow through mass transfer when there are instabilities of the disc. However, these disc 
instabilities are thought to result in the formation of the Galactic bar \citep{De-Lucia:2008}, whereas 
we are mainly interested in the properties of the Galactic spheroid in the Solar neighborhood area. 
Nonetheless, the accreted spheroid also contains stars that are situated in the Galactic bulge region. 
We define this region as the innermost 3 kpc of the spheroid, a definition that was also used by \citet{Cooper:2010}.
In section~\ref{sec:2.3}, we explain how we separate the bulge part and the halo part of the stellar spheroid,
to be able to focus on halo WDs in the Solar neighborhood area. But first, we summarize how stellar spheroids 
evolve in our model in sections~\ref{sec:2.1}, \ref{sec:2.2} and \ref{sec:2.3}.

\begin{figure*}
  \centering 
  \resizebox{\hsize}{!}{\includegraphics{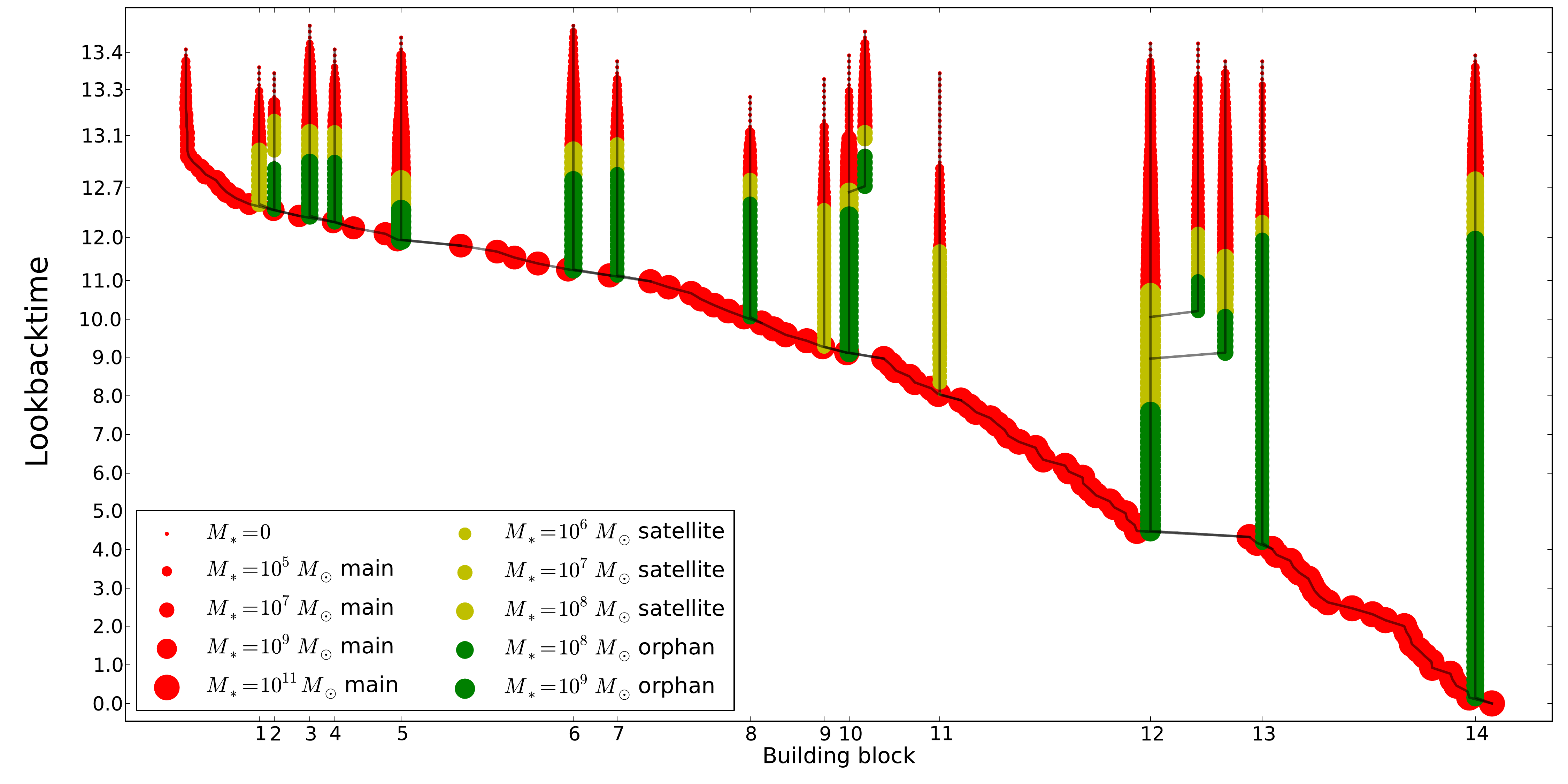}} 
  \caption{The galaxy merger tree of Aquarius halo A-4, showing only those objects that contribute at least 0.1\% to the total stellar mass of the
  accreted spheroid (in this example this corresponds to galaxies with a minimum stellar mass of $4.5 \cdot 10^6 M_\odot$). 
  Only if a building block of a building block itself has a stellar mass above this threshold it is also shown.
  Building block 12 is the largest progenitor building block, contributing 45\% of the accreted spheroids' stellar
  mass, followed by building blocks 5, 10 and 6, which contribute respectively 31\%, 10\% and 4.9\%.
  }
  \label{fig:01}
\end{figure*}

\subsection{The semi-analytic galaxy formation model}\label{sec:2.1}
The semi-analytic techniques that we use in our galaxy formation model originate in Munich
\citep{Kauffmann:1999,Springel:2001,De-Lucia:2004}, and were subsequently updated by many other authors
\citep{Croton:2006,De-Lucia:2007,De-Lucia:2008,Li:2009,Li:2010,Starkenburg:2013}, including some implemented 
in Groningen. Hence, we refer to this model as the Munich-Groningen semi-analytic
galaxy formation model. Note, that the ‘ejection model’ described by \citet{Li:2010} was also used by
\citet{De-Lucia:2014}. It is beyond the scope of this paper to summarize all the physical prescriptions
of this model \citep[as is done, eg. by][]{Li:2010}. Instead, we will focus on the evolution of the accreted
spheroid after we apply our model to five of the six high-resolution dark matter halo simulations of the Aquarius project
\citep{Springel:2008}.

The Aquarius dark matter haloes were selected from a lower resolution parent simulation 
because they had roughly Milky Way mass and no massive close neighbor at redshift 0.
The five dark matter haloes that we use, labelled A$-$E\footnote{Aquarius halo F was not used, because it experienced 
a recent significant merger and is therefore considered to be less similar to the Milky Way than the other five haloes.}, 
were simulated at 5 different resolution levels. The lowest resolution simulations, 
in which the particles had a mass of a few million $M_\odot$, are labelled by the number 5,
with lower numbers for increasingly higher resolution simulations, up to a few thousand $M_\odot$ per particle 
for resolution level 1. Only Aquarius halo A was run at the highest resolution level, but all haloes were
simulated at resolution level 2, corresponding to $\sim 200$ million particles per halo, or $\sim 10^4 M_\odot$
per particle. This is the resolution level that we use throughout this paper. 
The $\Lambda$CDM cosmological parameters in Aquarius are $\Omega_m = 0.25$, 
$\Omega_\Lambda = 0.75$, $\sigma_8 = 0.9$, $n_s = 1$, $h = 0.73$ and $H_0 = 100 h$ km s$^{-1}$ Mpc$^{-1}$.
The \textsc{subfind} algorithm \citep{Springel:2001} was used on the Aquarius simulations to construct a dark matter merger tree 
for a Milky Way-mass galaxy and its substructure, which can be used as a backbone to construct a galaxy merger tree. 
From this, we can determine if and when galaxies merge with other galaxies, following prescriptions for stellar stripping 
and tidal disruption of satellite galaxies \citep{Starkenburg:2013}.

The merger tree of the the modelled Milky Way in Aquarius halo A-4 is plotted in Figure~\ref{fig:01}. This is a slightly
lower resolution simulation than we use throughout the rest of this paper, but it suits the visualization purpose of this
Figure. The number of significant building blocks and their relative mass contributions to the fully accreted spheroid of
Aquarius halo~A is almost identical to that in resolution level 2. Time runs downwards in Figure~\ref{fig:01} 
and each circle denotes a galaxy in a different time step. The size of the circle indicates the
stellar mass of the galaxy. The building blocks of the Milky Way are shown as straight lines from the top
of the diagram (early times) until they merge with the main branch of the merger tree, which is the only line that is not running
vertically straight\footnote{Although some building blocks merged with the main branch less than a few Gyr ago, 
they stopped forming stars much earlier.}. 
Each building block is given a number, this is indicated on the horizontal axis. 
The four major building blocks of the stellar halo in this case collectively contribute more than 90\% of its stellar mass.

In merging with the Milky Way, each building block undergoes three phases. At first, it is a galaxy on its own 
in a dark matter halo. During this phase, the building block is visualized as a red circle in Figure~\ref{fig:01}.
As soon as its dark matter halo becomes a subhalo of a larger halo, the galaxy is called a satellite galaxy and 
the circles' colour changes to yellow. Once the dark matter halo is tidally stripped below the \textsc{subfind} resolution 
limit of 20 particles, it is no longer possible to identify its dark matter subhalo.
Because they have ``lost'' their dark matter halo, we call these galaxies orphans, and the corresponding circles
are coloured green. 

The semi-analytic model assumes that stars above 0.8~$M_\odot$ die instantaneously 
and that those below 0.8~$M_\odot$ live forever. This is also known as the instantaneous recycling 
approximation (IRA). Throughout this paper, the metallicity values predicted by our model are expressed 
as $\log[Z_\mathrm{stars}/Z_\odot]$, with $Z_\mathrm{stars}$ the ratio of mass in metals 
over the total mass in stars, and $Z_\odot = 0.02$ the Solar metallicity. 

\subsection{Spheroid star formation}~\label{sec:2.2}
\begin{figure}
  \centering 
  \resizebox{\hsize}{!}{\includegraphics{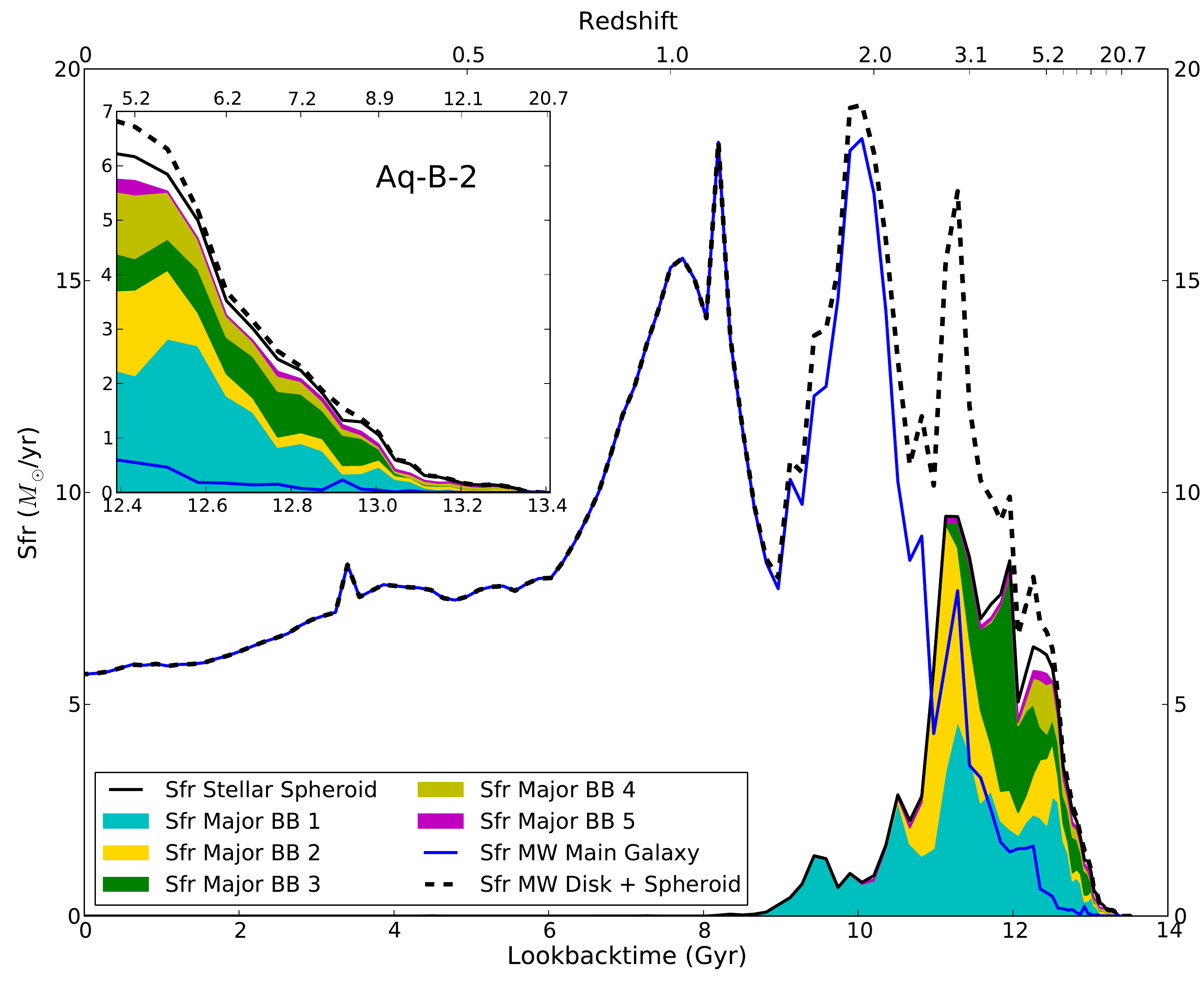}} 
  \caption{Star formation rate of the Milky Way mass galaxy in Aquarius halo B-2 (blue solid line) 
  and the star formation rate of its stellar spheroid (black solid line) as a function of time.
 Contributions from the five most massive building blocks are indicated by different colours (see legend). The 
 black dashed line indicates the complete SFH of the simulated galaxy at $z=0$, i.e. the sum of the blue and the
 black solid line. The corresponding redshift at each time is labelled on the top axis. 
 At early times, i.e. the first Gyr of star formation, which is shown in 
 the zoom-in panel, the star formation in some of the building blocks was much higher than that in the disc of the main galaxy.}
  \label{fig:02}
\end{figure}

A stellar halo of Milky Way mass is known to have only a few main progenitor galaxies
\citep[eg.]{Helmi:2002,Helmi:2003,Font:2006,Cooper:2010,Gomez:2013}.
We show the SFR in Aquarius halo B-2 as an example of the building blocks' contribution to the total star formation history 
of a Milky Way mass galaxy in Figure~\ref{fig:02} (for more details, see \citet{van-Oirschot:2017}, hereafter Paper~II).
With a blue solid line, the SFR in the disc is visualized, and with a black solid line 
that in the discs of building block galaxies, collectively forming the SFR of the modelled galaxy's spheroid. The dashed black line
is the sum of these two lines. With five different colours, contributions from the SFRs of the five most massive building 
blocks are visualized. As can be clearly seen from this figure, they collectively constitute almost the entire SFR of
the spheroid. In section~\ref{sec:4} we assume that stellar halo in the Solar neighborhood is built up entirely 
by four building blocks. This is in agreement with the simulations of streams in the Aquarius stellar haloes 
by \citet{Gomez:2013}, who used a particle tagging technique to investigate the Solar neighborhood sphere of 
the Aquarius stellar haloes with the \textsc{galform} semi-analytic galaxy formation model
\citep[see also][]{Cooper:2010}.

\subsection{The initial mass function} \label{sec:2.3}
The Munich-Groningen semi-analytic galaxy formation model assumes a \citet{Chabrier:2003} IMF.
As explained in Appendix~\ref{Ap:A}, the IRA applied to this IMF is equivalent to returning immediately
43\% of the initial stellar mass to the interstellar medium (ISM). However, as we also show in Appendix~\ref{Ap:A}, 
the return factor is a function of time (and of metallicity, to lesser extent). The value 0.43 is only reached after 13.5 Gyr, 
thus by making the IRA, our semi-analytic model over-estimates the amount of mass that is returned to the ISM at earlier times. 
We neglect this underestimation of the present-day mass that is locked up in halo stars, 
but we correct for the fact that stars have finite stellar lifetimes by evolving the initial stellar population 
with the binary population synthesis code SeBa. The details of our binary population synthesis model are set out 
in section~\ref{sec:3}.

It is not known if the Chabrier IMF is still valid at high redshifts, when the progenitors of the oldest WDs were born. 
Several authors have investigated top-heavy variants of the IMF \citep[eg.][]{Adams:1996,Chabrier:1996,Komiya:2007,Suda:2013},
initially to investigate if white dwarfs could contribute a significant fraction to the dark matter budget of the Galactic
spheroid, and later to explain the origin of carbon-enhanced metal-poor stars. In Paper~I, it was explored if the top-heavy 
IMF of \citet{Suda:2013} could be the high redshift form of the IMF, by comparing simulated halo WD luminosity
functions (WDLFs) with the observed one by \citet[hereafter RH11]{Rowell:2011}, derived from selected halo WDs in the
SuperCOSMOS Sky Survey. It was found that the number density of halo
WDs was too low to assume a top-heavy IMF, and that the \citet{Kroupa:1993} or \citet{Salpeter:1955} IMF result in halo WD 
number densities that match the observations better. 
We show in Appendix~\ref{Ap:A} that the \citet{Chabrier:2003} IMF is already more top-heavy than the Kroupa IMF, 
when it is normalized to equal the amount of stars with a mass below 0.8~$M_\odot$ for the Kroupa IMF. 
Because of the results of Paper~I, we therefore do not investigate more top-heavy alternatives of the \citet{Chabrier:2003} 
IMF \citep[eg.]{Chabrier:1996,Chabrier:1999} is this work.

\subsection{Selecting halo stars from the accreted spheroid} \label{sec:2.4}
As input for our population study of halo WDs, we use so-called age-metallicity maps. 
These show the SFR distributed over bins of age and metallicity. 
For the six Aquarius accreted stellar spheroids, the age-metallicity maps are shown in Figure~3 of Paper~II.
Halo WDs can only be observed in the Solar neighborhood (out to a distance of $\sim 2.5$~kpc with the Gaia
satellite, see Paper~I). 
Because we do not follow the trajectories of the individual particles that denote the building blocks 
\citep[as e.g. done by][]{Cooper:2010} we have to decompose the age-metallicity maps of the accreted spheroids
into a bulge and a halo part\footnote{We cannot use the publicly available results of \citet{Lowing:2015}, 
because they did not model binary stars and did not make WD tags.}. 
Making use of the observed metallicity difference between the bulge and the halo,
we select ``halo'' stars from the total accreted spheroid by scaling the metallicity distribution function (MDF) 
to the observed one.

We impose the single Gaussian fit to the observed photometric MDF of the stellar halo by \citet{An:2013}\footnote{
We decided not to use the two-component fit to the MDF that was determined by 
\citet{An:2013} to explore the possibility that there are two stellar halo populations, because the lowest metallicity population 
of halo stars is underrepresented in our model, as was already concluded from comparing the Aquarius accreted spheroid MDFs to 
observed MDFs of the stellar halo in Paper~II.}: 
$\mu_{[\mathrm{Fe}/\mathrm{H}]} = -1.55$, $\sigma_{[\mathrm{Fe}/\mathrm{H}]} = 0.43$. 
We use the MDF that was constructed from observations in the co-added catalog in 
SDSS Stripe 82 \citep{Annis:2014}. 
The stars that were selected from SDSS Stripe 82 by \citet{An:2013} are at heliocentric distances of 5$-$8 kpc,
thus this observed MDF is not necessarily the same as the halo MDF in the $\sim 2.5$~kpc radius sphere 
around the Sun that we refer to as the Solar neighborhood.
However, we consider it sufficient to use as a proxy to distinguish the halo part 
of our accreted spheroids' MDFs from the bulge part in our models.
The single Gaussian that we used was expressed in terms of [Fe/H], whereas the metallicity values in our model
can better be thought of as predictions of [$\alpha$/H], because of the IRA. Using an average [$\alpha$/Fe] value of 0.3~dex 
for the $\alpha$-rich (canonical) halo \citep{Hawkins:2015}, we added this to the single Gaussian MDF to arrive at 
$\mu_{[\alpha/\mathrm{H}]} = -1.25$ ($\sigma_{[\alpha/\mathrm{H}]} = \sigma_{[\mathrm{Fe}/\mathrm{H}]} = 0.43$). 

\begin{figure*}
  \centering 
  \resizebox{0.78\hsize}{!}{\includegraphics{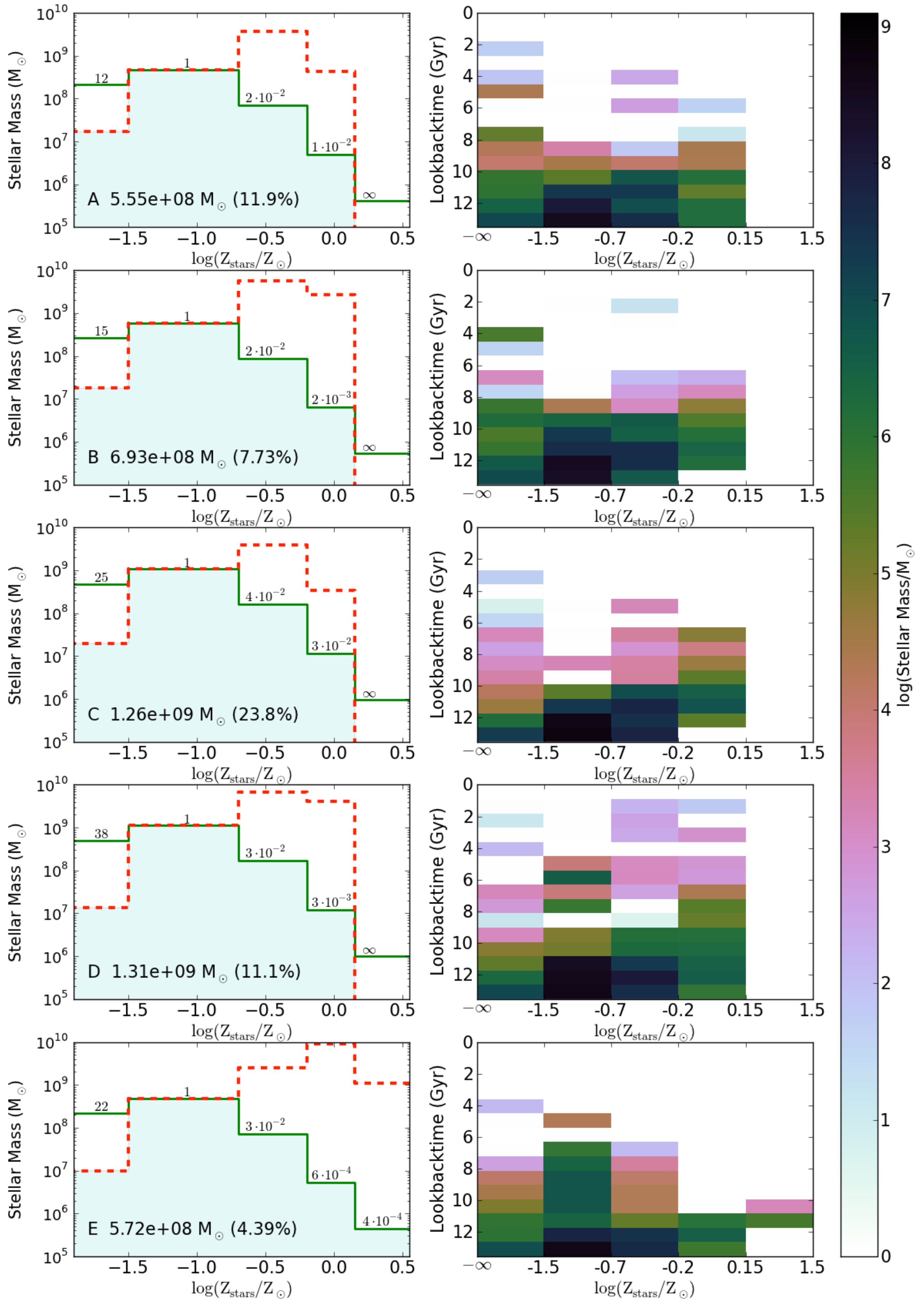}} 
  \caption{Left-hand side panels: MDF of the stellar halo in the Solar neighborhood,
  based on a single Gaussian fit to the observed photometric metallicity distribution (green solid lines) subtracted from 
  the co-added catalog in SDSS Stripe 82 \citep{An:2013} compared with the spheroid MDFs in our semi-analytical model 
  of galaxy formation combined with the Aquarius dark matter simulations (red dashed lines), for haloes A$-$E from top to bottom. 
  Here, 0.3 dex was added to the [Fe/H] values of the observed MDF to compare them with our model's $\log(\mathrm{Z}_\mathrm{stars}/\mathrm{Z}_\odot)$ values 
  \citep[based on an estimation of the $\lbrack\alpha$/Fe$\rbrack$ value for the $\alpha$-rich (canonical) halo by][]{Hawkins:2015}, since 
  the metallicity values of our model can better be compared with [$\alpha$/H] than with [Fe/H].
  The numbers written on top of each bin of the observed MDF indicate the discrepancy between our model and the observed value
  (see text for details). The bin with $\log(\mathrm{Z}_\mathrm{stars}/\mathrm{Z}_\odot)$ between
  $-1.5$ and $-0.7$ was used for the normalization of the observed MDF.
  The shaded area indicates the model MDF that we use as input for this population synthesis study of halo stars.
  The text in this shaded area indicates the halo ID, the total stellar halo mass and the 
  percentage of the total accreted stellar spheroid mass that we assume to be in halo stars. 
  Right-hand side panels: Age-metallicity maps ($\log(\mathrm{Z}_\mathrm{stars}/\mathrm{Z}_\odot)$) corresponding to the assumed stellar halo MDFs 
  in the left-hand side panels, again for haloes A$-$E from top to bottom. The colour map represents the stellar mass ($M_\odot$) per bin, 
  on a logarithmic scale. Note the non-linear horizontal axis corresponding to the different sizes of the metallicity bins. 
  The choice for this binning is explained in  section~\ref{sec:3}.
  }
  \label{fig:04}
\end{figure*}

The MDFs of the accreted stellar spheroids in Aquarius haloes are shown with dashed red lines 
in the left-hand side panels of Figure~\ref{fig:04}, for haloes A$-$E from top to bottom.
In each panel, the green solid line indicates the number of stars in each metallicity bin according to the (shifted) single
Gaussian fit to the observed MDF by \citet{An:2013}, where the observations were normalized to the number of stars in the
$-1.5 \leq \log(\mathrm{Z}_\mathrm{stars}/\mathrm{Z}_\odot) \leq -0.7$ bin. The numbers written on top of each bin of the observed MDF 
indicate how much the red dashed line should be scaled up (when $> 1$) or down (when $< 1$) in that bin to match it with the green 
solid line\footnote{Since in haloes A$-$D, the accreted spheroids were not found to have any stars with $\log[Z_\mathrm{stars}/Z_\odot]$ ([$\alpha$/H]) values
above 0.15, these bins are labelled with the $\infty$-sign.}.

Although we underestimate the number of halo stars with the lowest metallicities ($\log(\mathrm{Z}_\mathrm{stars}/\mathrm{Z}_\odot) \lesssim -2$) 
in our model, we cannot increase this number, because that would imply creating extra stars. We can however reduce the number of
high metallicity halo stars, by ``putting them away'' in the bulge. We thus interpret all low-metallicity accreted spheroid stars 
as halo stars, and a large fraction of the high-metallicity stars as bulge stars. When we lower the number of stars in a metallicity bin, 
we do that with the same factor for all ages. The resulting input MDF is the shaded area 
in each of the panels in the left-hand side of Figure~\ref{fig:04}. In the right-hand side panels, 
we show the corresponding ages of the remaining stars in each metallicity 
bin. The colour map indicates the stellar mass on a logarithmic scale.

\section{Binary population synthesis}~\label{sec:3}

To model the evolution of binary WDs, we use the population synthesis code SeBa \citep{Portegies-Zwart:1996, Nelemans:2001a, Toonen:2012, Toonen:2013},
which was also used in Paper~I. In SeBa, ZAMS single and binary stars are generated with a Monte Carlo-method.
On most of the initial distributions, we make the same assumptions as were made in Paper~I, i.e.:
\begin{itemize}
\item Binary primaries are drawn from the same IMF as single stars
\item Flat mass ratio distribution over the full range between 0 and 1, thus for secondaries $m_\mathrm{low} = 0$
and $m_\mathrm{high} = m_\mathrm{primary}$.
\item Initial separation ($a$): flat in $\log a$ (\"Opik's law) between 1 R$_\odot$ and $10^6$ R$_\odot$ \citep{Abt:1983},
provided that the stars do not fill their Roche lobe.
\item Initial eccentricity ($e$): chosen from the thermal distribution $\Xi(e) = 2e$
between 0 and 1 as proposed by \citet{Heggie:1975} and \citet{Duquennoy:1991}.
\end{itemize}
However, instead of using \citet{Kroupa:1993} IMF as standard, we choose the \citet{Chabrier:2003} IMF in this paper
to match the initial conditions of our population of binary stars as much as possible to those in the Munich-Groningen
semi-analytic galaxy formation model (see also section~\ref{sec:2.3}).

\begin{figure}
  \centering 
  \resizebox{\hsize}{!}{\includegraphics{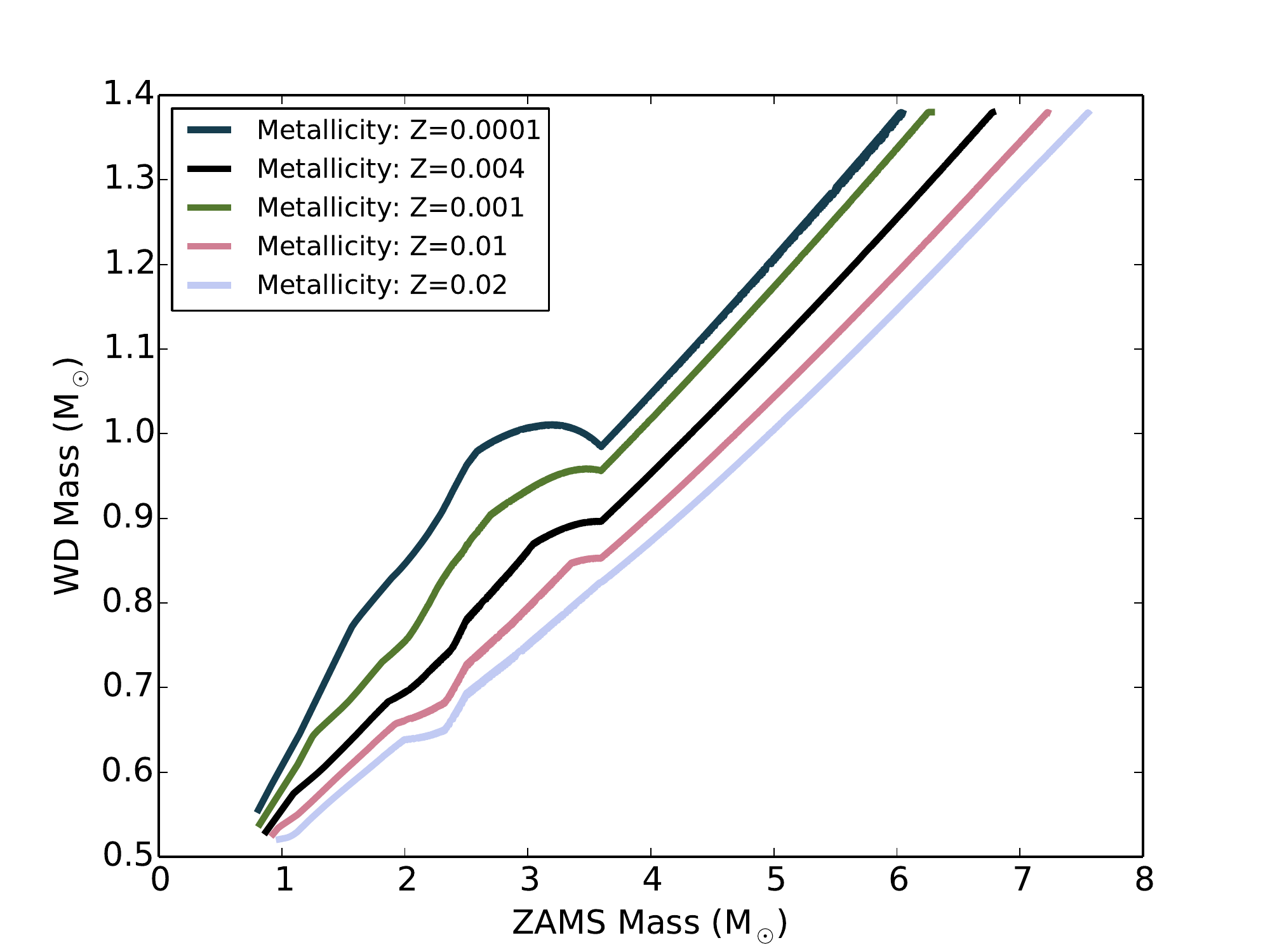}} 
  \caption{Initial to final mass relation (IFMR) for WDs with the five different metallicities used in this work.
  Based on the analytic formulae in \citet{Hurley:2000} and similar to their Figure~18.
  Note that with this choice of metallicity values there is an approximately equal distance between the five lines,
  so by simulating a stellar population in which the stars have one of these five metallicity values, the effect of 
  metallicity on the IFMR is fully covered.}
  \label{fig:03}
\end{figure}

We evolve a population of halo stars in a region of $\sim 3$~kpc around the Sun 
(see Paper~I for more details\footnote{Note that the boundary condition given in equation
A.11 of Paper~I contains a small error: $\nicefrac{\pi}{2}$ should be $\pi$.}).
This population is modelled with five different metallicities: 
$Z=0.02$, $Z=0.01$, $Z=0.004$, $Z=0.001$ and $Z=0.0001$. 
The choice for these five metallicity values was motivated by our aim to cover as much as possible the effect of metallicity on 
the initial-to-final-mass relation for WDs (IFMR), see Figure~\ref{fig:03}.
These metallicities correspond with the bins we use in the semi-analytic galaxy formation model (Figure~\ref{fig:04}) when correcting
for the fact that the semi-analytic model gives [$\alpha$/H] that are 0.3~dex higher than [Fe/H]\footnote{The lowest
metallicity bin is chosen to extend to $-\infty$, in order to also include stars with zero metallicity. 
These (still) exist in our model because we neglect any kind of pre-enrichment from Population III stars.}.

The evolution of the stars is followed to the point where they become WDs, neutron stars, or black holes. A binary system is followed until 
the end-time of the simulation, considering conservative mass transfer, mass transfer through stellar winds or dynamically unstable mass transfer 
in a common envelope in each time step with approximate recipes \citep[see][and references therein]{Toonen:2013}.
Also angular momentum loss due to gravitational radiation, non-conservative mass transfer or magnetic braking is taken into account. 
To follow the cooling of the WDs, we use a separate method, explained in the next subsection.

\subsection{White dwarf cooling and Gaia magnitudes}
\begin{figure}
  \centering 
  \resizebox{\hsize}{!}{\includegraphics{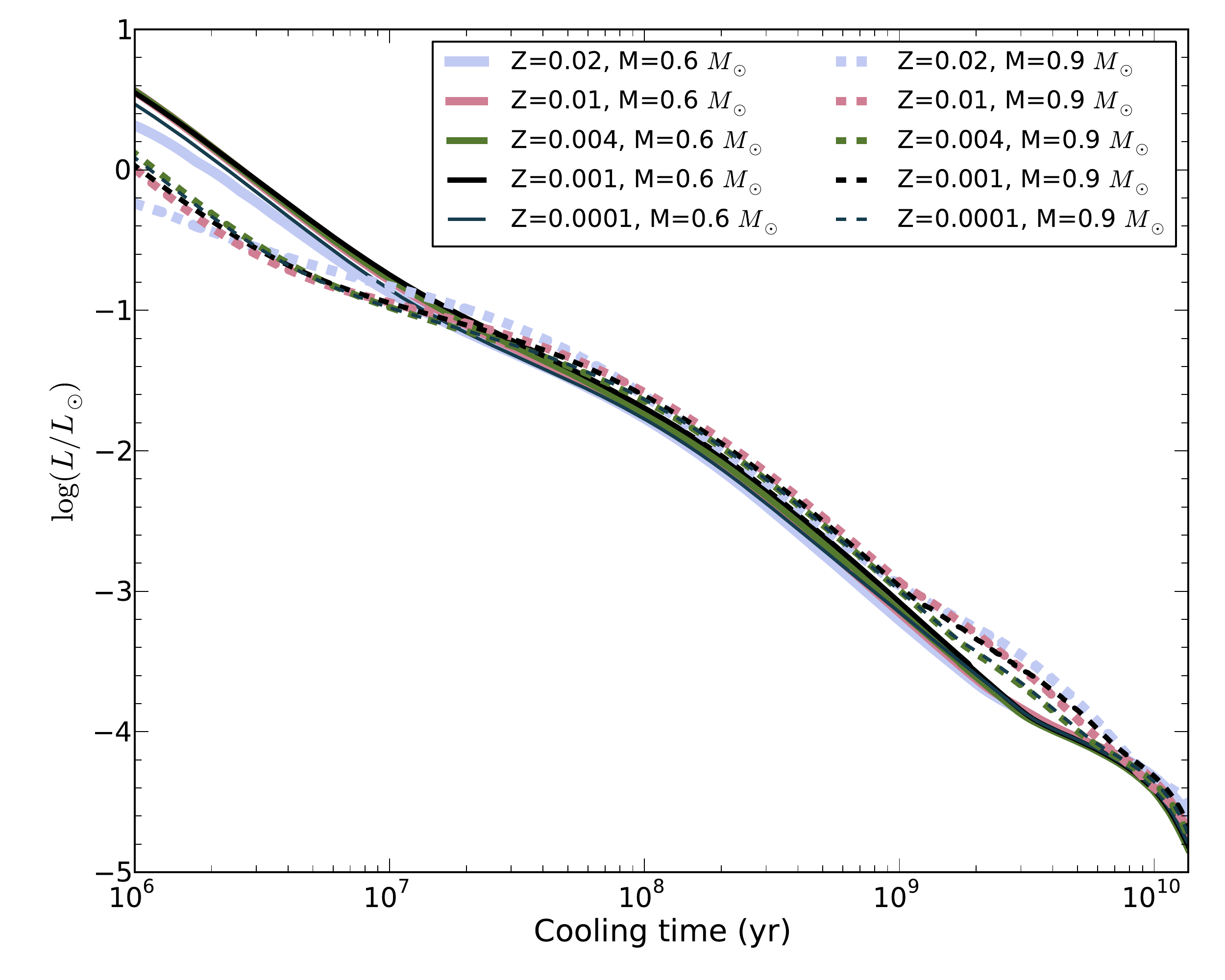}} 
  \caption{White dwarf luminosity as a function of age, for WDs having progenitor stars with 5 different metallicities, 
  for two different masses. Interpolation was used on cooling tracks calculated by several authors: \citet{Salaris:2010} for
  Z=0.02, \citet{Renedo:2010} for Z=0.01, \cite{Romero:2015} for Z=0.004 and Z=0.001, and \citet{Althaus:2015} for Z=0.0001.  }
  \label{fig:05}
\end{figure}

We use the recent work on the cooling of carbon-oxygen (CO) WDs with low metallicity progenitor stars \citep{Renedo:2010,Althaus:2015,Romero:2015}
to calculate the present day luminosities and temperatures of our simulated halo WDs with sub-Solar metallicity. For those
with Solar metallicity, we use the cooling tracks that were made publicly available by \citet{Salaris:2010}. As in Paper~I, we interpolate
and extrapolate the available cooling tracks in mass and/or cooling time, to cover the whole parameter space that is sampled by our population synthesis code. 
The resulting cooling tracks for two different WD masses at five different metallicities are compared in Figure~\ref{fig:05}.
Although the effect of a different progenitor metallicity on the WD cooling is small, we still take it into account for 
WDs with a CO core.

Unfortunately, there were no cooling tracks for helium (He) core and oxygen-neon (ONe) core WDs with progenitors having 
a range of low metallicity values available to us for this study. For WDs with these core types, we therefore used the same cooling 
tracks for all metallicities \citep{Althaus:2007,Althaus:2013}. As in Paper~I, the extrapolation in mass is done such that for the WDs 
with masses lower than the least massive WD for which a cooling track is still available in the literature, the same cooling
is assumed as for the lowest mass WD that is still available. The same extrapolation is chosen on the high mass end.
The available mass ranges, as well as those in our simulation, are listed in Table~\ref{tab:01}. 
We extrapolate any cooling tracks that do not span the full age of the Universe. 
At the faint end of the cooling track we do this by assuming \citet{Mestel:1952} cooling. 
At the bright end, we keep the earliest given value constant to zero cooling time.

\begin{table}
\caption{White dwarf mass ranges in our simulation (S) and those for which cooling tracks are available in the literature (L).}
\begin{tabular}{lccccc}
\toprule 
							& $Z$=0.0001 & $Z$=0.001 & $Z$=0.004 & $Z$=0.01 & $Z$=0.02 \\
\midrule
\multirow{2}{*}{He (S)} 	& 0.144	& 0.161	& 0.148	& 0.146	& 0.142	\\
							& 0.509 & 0.496 & 0.487 & 0.481 & 0.476\\
\midrule
\multirow{2}{*}{CO (S)}	& 0.330 & 0.330 & 0.330 & 0.330 & 0.330 \\
							& 1.38 & 1.29 & 1.33 & 1.35 & 1.38 \\
\midrule
\multirow{2}{*}{CO (L)} & 0.520 & 0.505 & 0.503 & 0.525 & 0.54 \\
						  & 0.826 & 0.863 & 0.817 & 0.934 & 1.20 \\
\bottomrule
\end{tabular}
\tablefoot{\small The mass range for He core WD cooling tracks that are available in the literature is $0.155-0.435$
(only available for metallicity $Z=0.01$). $V$ and $I$ magnitudes as a function of cooling time for He core WDs are only
available for WDs in the mass range $0.220-0.521$, whose progenitors have metallicity $Z=0.03$.
The mass range for ONe WD cooling tracks that are available in the literature is $1.06-1.28$ 
(only available for metallicity $Z=0.02$). The simulations yield ONe WDs in the mass range $1.10-1.38$ 
(this simulated mass range is the same for all metallicities).}
\label{tab:01}
\end{table}

\begin{figure}
  \centering 
  \resizebox{\hsize}{!}{\includegraphics{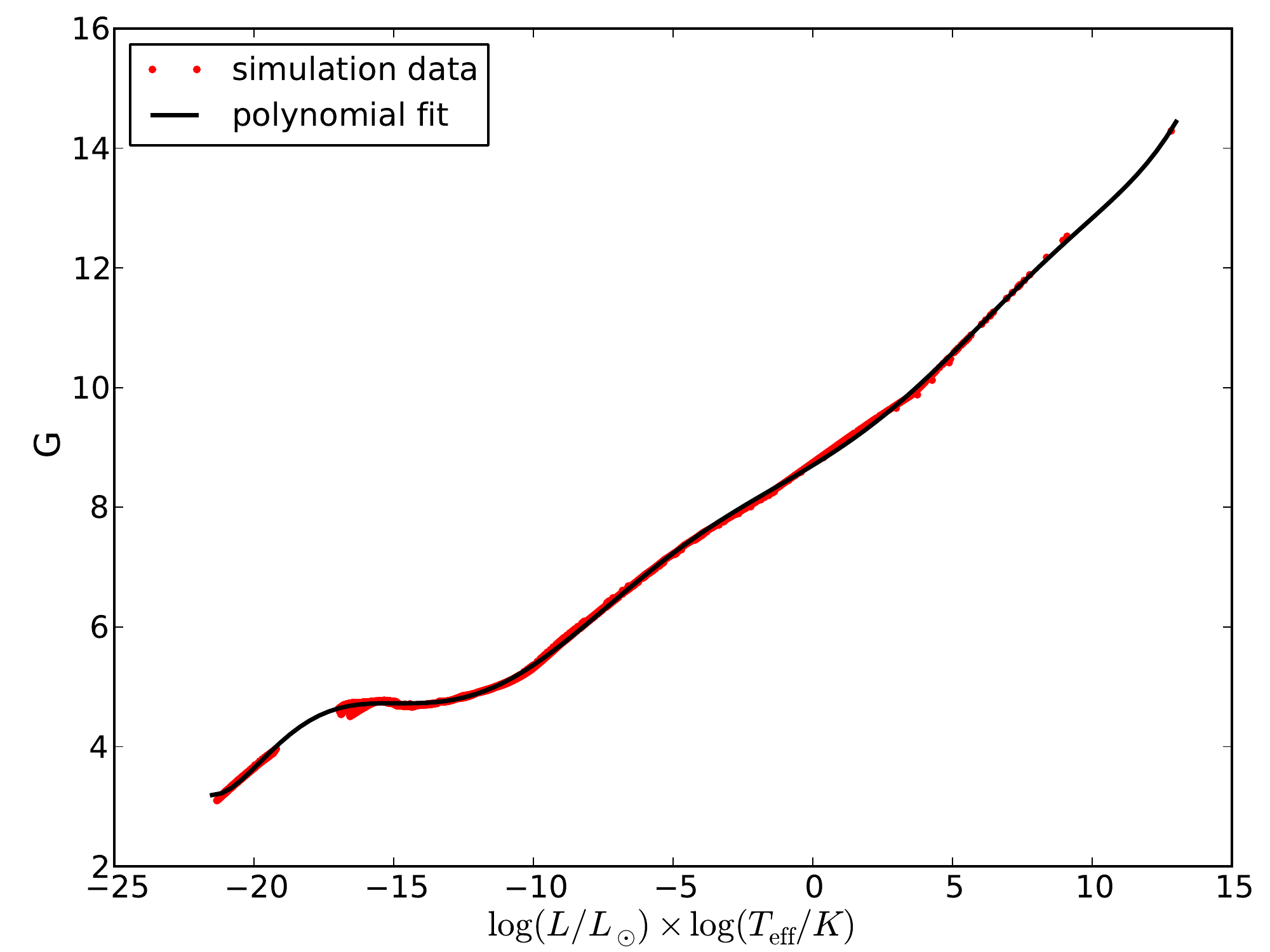}} 
  \caption{Gaia magnitude as a function of the product luminosity $\times$ temperature.
  The red poins is the simulation data of Paper~I. The black line is a polynomial fit to the data of degree 9, i.e.
  $G = a_0 x^9 + a_1 x^8 + \ldots + a_8 x + a_9$ with $x = \log (L/L_\odot) \times \log (T_\mathrm{eff}/K)$
  and function parameters $a_0 = -8.197\cdot 10^{-11}$, $a_1 = -6.837\cdot 10^{-10}$, $a_2 = 8.456\cdot 10^{-8}$, $a_3 = 7.256\cdot 10^{-7}$, 
  $a_4 =-2.347\cdot 10^{-5}$, $a_5 = -1.370\cdot 10^{-4}$, $a_6 = 2.451\cdot 10^{-3}$, $a_7 = 1.109\cdot 10^{-2}$, $a_8 = 2.866\cdot 10^{-1}$ and
  $a_9 = 8.701$.
 }
  \label{fig:06}
\end{figure}

We found that the Gaia magnitude can be directly determined from the luminosity and temperature of the WD
for CO and ONe WDs, rather than from synthetic colours and a colour transformation as done in Paper~I (see Figure~\ref{fig:06}).
For He core WDs, such a relation does not hold. For those, we apply the same method as in Paper~I.

To estimate by which amount the light coming from the WDs gets absorbed and reddened by interstellar dust before
it reaches the Gaia satellite or an observer on Earth, we assume that the dust follows the distribution 
\begin{equation}
P(z) \propto \mathrm{sech}^2(z/z_h),
\end{equation}
where $z_h$ is the scale height of the Galactic dust (assumed to be 120 pc) and $z$ 
the cartesian coordinate in the $z$-direction. As in Paper~I, we assume that the interstellar extinction between the observer 
and a star at a distance $d=\infty$ is given by the formula for $A_V(\infty)$ from \citet{Sandage:1972}, from which
follows that the $V$-band extinction between Gaia and a star at a distance $d$ with Galactic latitude $b = \arcsin (z/d)$,
\begin{equation}
A_V(d) =  A_V(\infty) \ \mathrm{tanh} \left(\frac{d\sin b}{z_h}\right). \label{Avd}
\end{equation}

\section{Halo WDs in the Solar neighborhood}\label{sec:4}
In this section, we investigate if the cosmological building block to building block variation is reflected in the present-day 
halo WD population, and if it is still possible to observationally distinguish
halo WDs originating from different building blocks of the Galactic halo.
Selecting four building blocks from each of the Aquarius stellar spheroids, scaled down in mass to disentangle stellar halo
from bulge stars, as explained in section~\ref{sec:2.4}, we present masses, luminosities and binary period distributions
of five cosmologically motivated stellar halo WD populations in the Solar neighborhood. 

Dividing the mass in each bin of a building block's age-metallicity map by the lock-up fraction of the semi-analytic model ($\alpha = 0.57$, see Appendix~\ref{Ap:B})
gives us the total initial mass in stars that was formed. The IMF dictates that 37.2\% of these stars will not evolve in 13.5 Gyr (see Appendix~\ref{Ap:A}). 
We thus know how much mass is contained in these so-called unevolved stars for each building block of our five simulated stellar haloes.
For each of the Aquarius haloes, we then choose four building blocks (after the modification of the accreted spheroid age-metallicity maps 
to stellar halo age-metallicity maps visualized in Figure~\ref{fig:04}) to represent the building blocks contributing to the stellar halo in the
solar neighborhood. The stars in these selected building blocks span multiple bins of the age-metallicity map, 
although the majority of stars is in the old and metal-poor bins.

The four building blocks of the stellar halo in the Solar neighborhood are selected such that they collectively 
have a MDF that follows the one we used in Figure~\ref{fig:04} to scale down the accreted spheroids' age-metallicity map
to one that only contains stars that contribute to the stellar halo. However, we do have some freedom in selecting which age bins 
contribute in the Solar neighborhood. We expect that the most massive building blocks of the stellar halo cover a volume 
that is larger than that of our simulation box, thus if such a building block is selected,
we assume that only a certain fraction its total stellar mass contributes to the Solar neighborhood.
The same fraction of stars is taken from all bins of its age-metallicity map,
to not change the age versus metallicity distribution of its stars. 
The total mass in unevolved stars in our simulation box is set to equal the amount estimated from the observed mass density 
in unevolved halo stars in the Solar neighborhood by \citet{Fuchs:1998} (see Appendix~A of paper~I).

By investigating the variety of building blocks of the Aquarius stellar spheroids, we found that the least massive building blocks 
have stars only in one or two bins of the age-metallicity map. Most of them are in the lower-left corner of the
age-metallicity map, where old and metal-poor stars are situated. To end up with only four building blocks contributing to the
Solar neighborhood and a MDF that follows the one we used in Figure~\ref{fig:04} we thus expect a selection of more
massive building blocks. Here, we aim to verify if it is possible to identify differences in the properties of
halo WDs due to their origin in different Galactic building blocks. Therefore, we select the building blocks
to contribute to the Solar neighborhood such that their overlap in the different bins of the age-metallicity map 
is as small as possible. One should keep this in mind when reading the remainder of this section. 
This is an optimistic scenario for finding halo white dwarfs in the Solar neighborhood with different
properties due to their origin in different Galactic building blocks in our model.

\begin{figure*}
  \centering 
  \resizebox{\hsize}{!}{\includegraphics{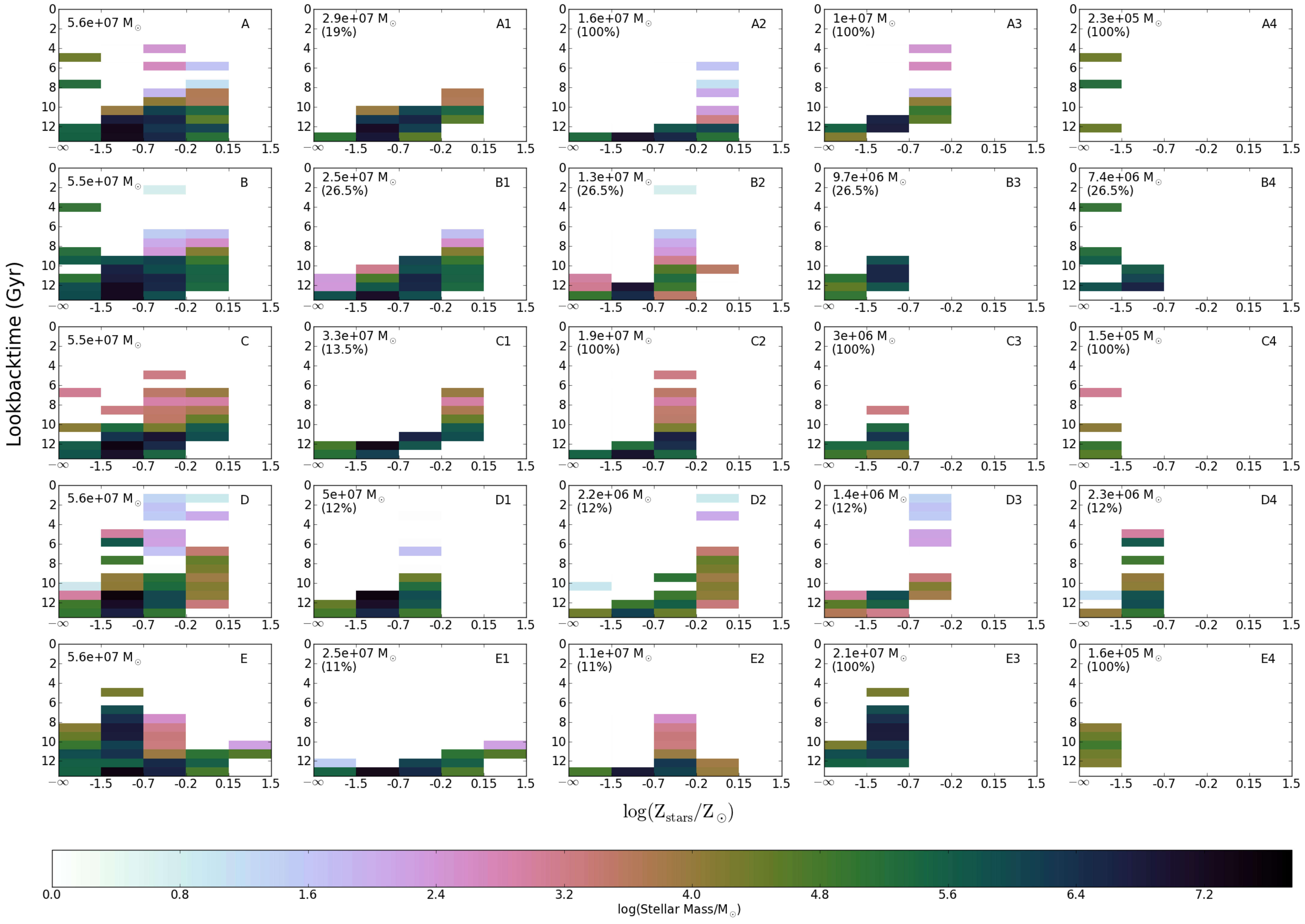}} 
  \caption{Age-metallicity maps of four selected building blocks from each halo. We have normalized the total mass in our Solar neighborhood volume 
  on the estimated mass density (in unevolved stars) by \citet{Fuchs:1998}. This results in a present day total stellar mass in halo stars in our 
  selected volume of $\sim 5.6\cdot 10^7 M_\odot$, as indicated in the upper left corner of the leftmost panels, which show the 
  summed age-metallicity maps of the four selected building blocks. The mass that each of the building blocks contribute to the Solar neighborhood 
  volumes is also indicated in the upper left corner of each of their panels, as is a percentage showing what fraction of the total stellar halo 
  building block (after our modifications to match it to the green lines in Figure~\ref{fig:04}) this mass corresponds to. 
  }
  \label{fig:11}
\vspace{0.5cm}
\centering	
	\resizebox{\hsize}{!}{\includegraphics{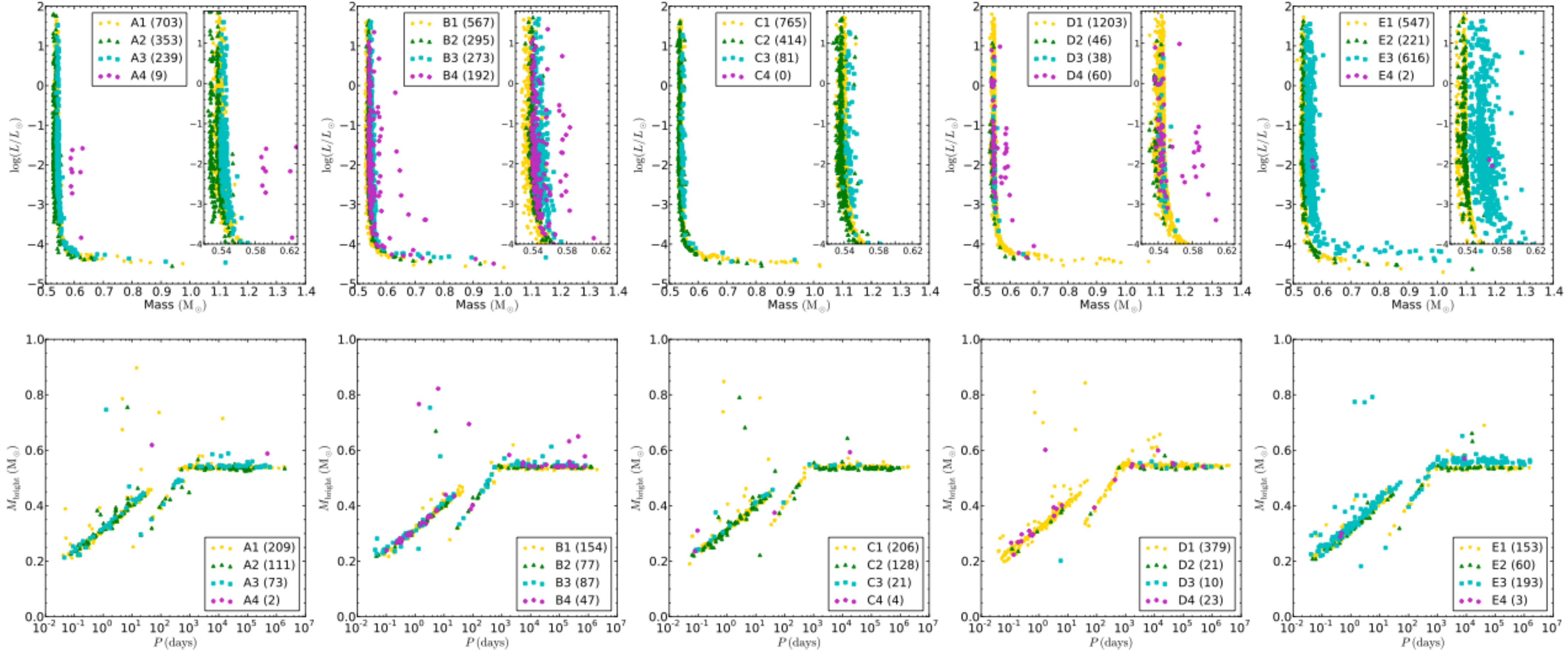}}
		\caption{Top panels: mass versus luminosity diagrams for the single WDs in the five Aquarius stellar halo populations 
		built using the age-metallicity maps of the four building blocks of each halo presented in Figure~\ref{fig:11}.
		In the upper left corner of each of panel we zoomed in on those WDs with masses between 0.52 and 0.63 $M_\odot$ and $\log (L/L_\odot) \geq -4$. 
		Bottom panels: period versus mass of the brightest WD of the unresolved binary WDs in these same simulated stellar haloes.
		}
		\label{fig:12}
\end{figure*}

In Figure~\ref{fig:11} we show the age-metallicity maps of the four selected building blocks, for Aquarius haloes A$-$E from top to bottom. The sum of the 
age-metallicity maps of the four building blocks is shown in the leftmost panels. When compared with the total age-metallicity maps
of our stellar spheroids (right-hand side panels of Figure~\ref{fig:04}) we see that most features of the total age-metallicity maps are covered
by these Solar neighborhood ones. The percentage of the total mass of that building block that we chose to be present in our
simulation box is shown in the upper left corner of each building block panel. In this corner the total mass of that age-metallicity map
is also shown (also in the leftmost panels).

With these four building blocks as input parameters for our binary population synthesis model, we made
mass versus luminosity diagrams for the single halo WDs with $G<20$ and period versus mass of the brightest WD 
of unresolved binary WDs  with $G<20$ in our simulations. These are shown in Figure~\ref{fig:12}.
The WDs of each building block are plotted with a separate colour and marker.
The numbers in between brackets in the legend indicate how many WDs (top panels) or unresolved binaries (bottom panels)
have $G<20$ and are plotted in the diagram. For building block C4 this equals 0 and also building blocks A4
and E4 contribute less than 10 single WDs to the stellar halo in the Solar neighborhood. This is because the
masses of these building blocks are so small that all WDs that are present in that building block at the present day
have $G\geq 20$.

The bottom panels of Figure~\ref{fig:12} show that there are no large differences between the simulated haloes including their
distinct building blocks in the period versus mass of the brightest WD in unresolved binary WD systems. All five diagrams look
more or less the same, and all building blocks cover the same areas in the diagram, although some naturally have more binary WDs
(with $G<20$) than others.

The top panels of Figure~\ref{fig:12} reveal that the mass versus luminosity diagrams of single halo WDs show slightly larger
differences between the simulated haloes and building blocks. The 9 WDs originating from building block A4 have clearly higher
masses than those that were born in the other three selected building blocks of Aquarius halo A, which can be understood from its
age-metallicity map (the top right panel of Figure~\ref{fig:11}). A large fraction of the stars in A4 is young and metal-poor,
thus based upon Figure~\ref{fig:08} we expect that many halo WDs from A4 are located to the right of the main curve in this diagram.
The same explanation holds for some WDs from building blocks B4 and D4. There are no large differences between the single halo WDs from 
the building blocks of simulated Aquarius halo C in this diagram. The selected building blocks of Aquarius stellar halo E result in 
a single halo WD population with a wide mass range in these panels, i.e approximately two times the width of the mass range of the single 
halo WD population in halo C. This is due to the many young stars in building block E3.

With standard spectroscopic techniques WD masses can be determined with an accuracy of $\sim$0.04~$M_\odot$ \citep{Kleinman:2013},
which would make it hard, though not impossible to identify some of the signatures described in the previous paragraph. With high-resolution
spectroscopy accuracies of $\sim$0.005~$M_\odot$ can be obtained \citep{Kalirai:2012}, which would make it much easier to identify these signatures.
However, there are two main issues that prevent us from drawing strong conclusions on this. Firstly, it is unclear whether the stellar halo of the Milky Way
in the Solar neighborhood is indeed composed out of building blocks which are as distinct from each other as those that we selected
in this work. We are comparing the haloes of only five Milky Way-like galaxies, that are dominated by a few objects, which makes this a stochastic result. 
Even for the optimistic scenario studied here, not in all five haloes we find distinct groups of single halo WDs in the mass versus luminosity diagram.
Halo C, for example, does not show this and for halo E there is no gap in the mass range spanned by the four building blocks, which 
makes it observationally impossible to disentangle contributions from the four building blocks. Secondly, it was shown in paper~I that a
WD that is the result of a merger between two WDs in a binary can end up in the mass versus luminosity diagram easily 0.1~$M_\odot$
left and right of the main curve, which (in the latter case) makes it indistinguishable from a single WD that was born in a separate 
building block.

We conclude that there are rather small differences between WDs in realistic cosmological building blocks.
In Appendix~\ref{Ap:C} we show what the maximum differences could be for haloes built from BBs that have wildly 
different ages and metallicities.

\section{The halo white dwarf luminosity function} \label{sec:5}

In this section, we will present the WDLFs for the five selected Aquarius stellar haloes from the previous section.
We will compare them to the observed halo WDLF by RH11 and also to the three best fit models of Paper~I.

In a recent paper, \citet{Cojocaru:2015} also investigated the halo WDLF. Although their work focusses on
single halo WDs, they also draw conclusions on the contributions from unresolved binaries. There are large differences
between their study and ours, the most important one being that they do not follow the binary evolution in detail, 
whereas we do. Therefore, our simulated WDs have different properties (mainly the Helium core WDs) which clearly results 
in a different luminosity function. \citet{Cojocaru:2015}'s statement that unresolved binaries are more often than single WDs 
found in the faintest luminosity bins, seems implausible together with our assumption that residual hydrogen burning in He-core WDs 
slows their evolutionary rate down to very low luminosities.
This was shown to be the case, at least for He-core WDs with high-metallicity progenitors, by \cite{Althaus:2009}\footnote{
The effect of a lower metallicity is expected to affect the lifetime previous to the WD stage and the thickness of the
hydrogen envelope. White dwarf stars with lower metallicity progenitors are found to have larger hydrogen envelopes 
\citep{Iben:1986,Miller-Bertolami:2013,Romero:2015} resulting in more residual H burning, which delays the WD cooling 
time even further. Overall, we find that the effect of progenitor metallicity on the WD cooling is not very
large, at least for CO WDs, for which cooling curves for WDs with different metallicity progenitors were 
available to us (see Figure~\ref{fig:05}) and are used in this paper.}.
It was shown in Paper~I that unresolved binaries mainly contribute to the halo WDLF at the bright end. In fact, $\sim$50\% of the stars 
contributing to the brightest luminosity bins of the halo WDLF ($M_\mathrm{bol}\lesssim 4$) are unresolved binary pairs.

The effective volume technique used by RH11 results in an unbiased luminosity function that can directly be compared 
to model predictions. Therefore, no series of selection criteria has to be applied to any complete mock database of halo WDs
before comparing it with their observational sample, although \citet{Cojocaru:2015} claim otherwise.
However, one should apply a correction for incompleteness in the survey of RH11. As we already explained in a footnote on page~10, 
we apply a correction factor of 0.74 to our model lines to compare them with the RH11 WDLF in this work.

\begin{figure}
  \centering 
  \resizebox{\hsize}{!}{\includegraphics{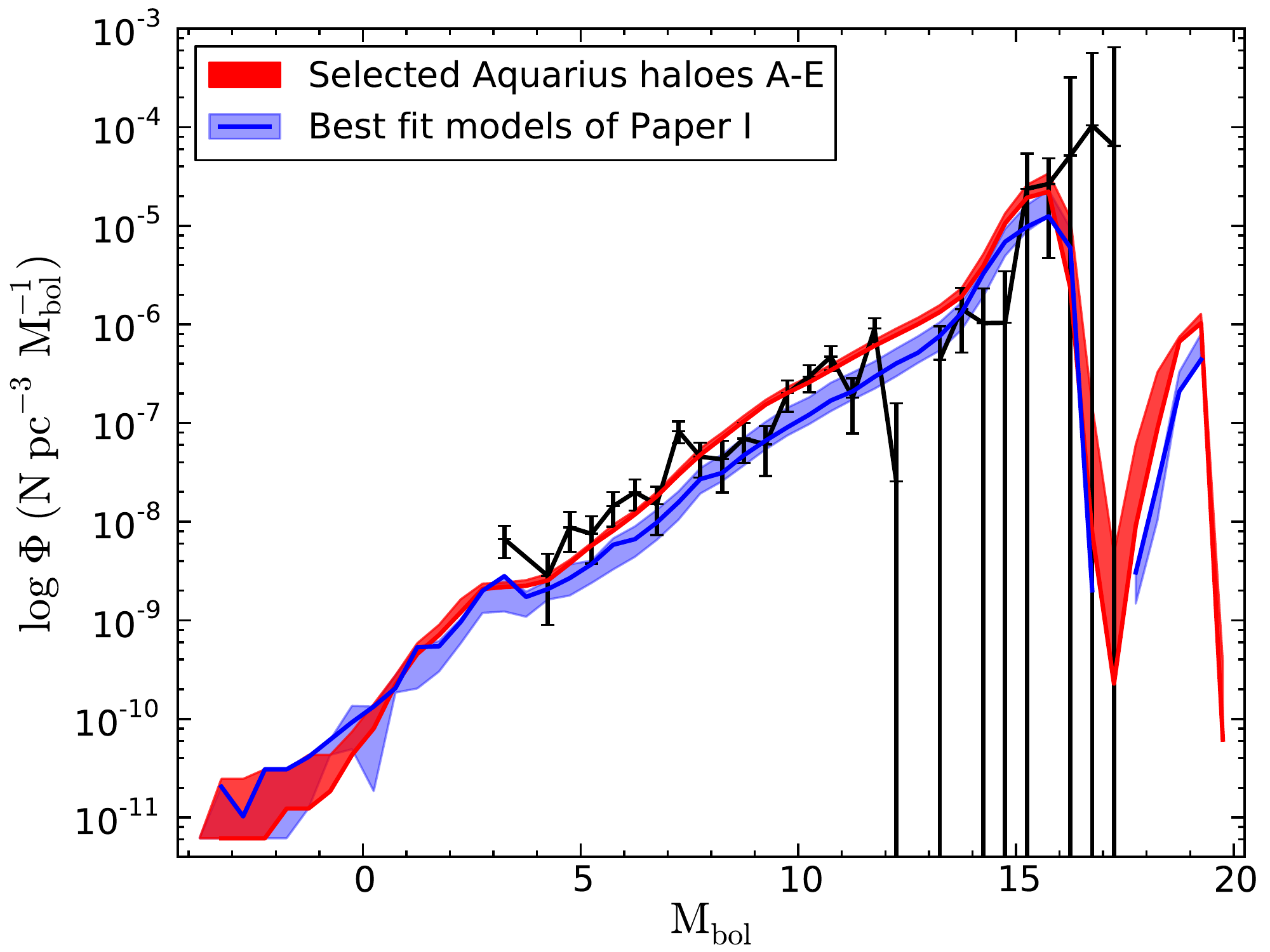}} 
  \caption{Halo white dwarf luminosity function for the five selected Aquarius stellar haloes
  (red band) compared to the observed one by RH11 (black line with errorbars). Also the three 
  best-fit models of Paper~I are shown with a blue band (see text for details). }
  \label{fig:13}
\end{figure}

The halo WD populations from the five selected Aquarius stellar halo WDs in the Solar neighborhood result five in halo
WDLFs that are very similar to each other. They are plotted as a single red band in Figure~\ref{fig:13}. The thickness of the band
indicates the spread in the five models, since the upper and lower boundary of the band indicate the maximum and minimum value of the WDLF
in the corresponding bin. With a black line with errorbars, RH11s observed halo WDLF is shown. The reduced $\chi^2$ values for the five
different Aquarius stellar halo selections are 3.4, 3.5, 3.3, 4.1 and 4.6 for haloes A$-$E respectively.
The fact that these five models are so similar is not surprising, given that the stellar haloes from which the four major building blocks
were selected all were modified to follow the same MDF, and normalized to observed local halo mass density in unevolved stars
\citep[][see Appendix~\ref{Ap:B}]{Fuchs:1998}. We again stress that it is remarkable that we find such an agreement with the
observed WDLF with this normalization, as we also found in the bottom right panel of Figure~\ref{fig:14} (see also Figure~4 of Paper~I).
Most other authors, including \citet{Cojocaru:2015}, simply normalize their theoretical WDLF to the observed one.

For comparison, the WDLFs predicted by the three best-fit models of Paper~I are shown with a light blue band in Figure~\ref{fig:13}. 
The blue line in this band corresponds to the 100\% binaries model (Kroupa IMF). For most bins, this line is in between the 
50\% binaries line with Kroupa IMF (upper boundary of the blue band) and Salpeter IMF (lower boundary of the blue band).
Since the correction factor for incompleteness that we apply in this work is slightly different from the one that was applied in Paper~I
(0.74 instead of 0.45), we find that the 100\% binaries model of Paper~I, with a reduced $\chi^2$ value of 2.2, actually fits the RH11 WDLF 
slightly better than the standard model in Paper~I. For both this standard model (Kroupa IMF, 50\% binaries) and the model with a Salpeter IMF 
we now find a reduced $\chi^2$ value of 2.4. In Paper~I the effect of a different normalization on the reduced $\chi^2$ values 
was already investigated. Since the corrected correction factor 0.74/0.45 = 1.64 is close to the optimal multiplication factor to
obtain a minimum reduced $\chi^2$ value for the 100\% binaries model, it is not surprising that this model comes out best. The small rise 
of the WDLF in the brightest bin could be due to the contribution from unresolved binaries (see Paper~I), but due to the normalization that we chose,
our model lines are too low in this bin. 

The fact that the lines in the blue band have a lower reduced $\chi^2$ value than the ones in the red band is mainly due to the bad fit 
at the faint end of the WDLF, i.e. in the bins centered at 12.25, 14.25 and 14.75. Since we normalize our model lines to the corresponding 
present day mass in unevolved local halo stars, $\rho_0$, and the Chabrier IMF is slightly more top-heavy than the Kroupa IMF, this leads to many more stars that 
have evolved to WDs at the present day (see Figure~\ref{fig:A2}). A similarly bad fit was seen for the top-heavy IMF used in Paper~I. 
However, the estimated $\rho_0$ has a statistical uncertainty that we did not take into account here. \citet{Fuchs:1998} found that the most 
likely value of $\rho_0$ lies in the range 1.5 to 1$\cdot 10^{-4} M_\odot \ \mathrm{pc}^{-3}$, with the latter value, in their view, 
being a firm lower limit. If we use this latter value of $\rho_0$ we find WDLFs that are two thirds lower than these ones. 
The red band in Figure~\ref{fig:13} would be shifted to the current position of the blue band, and its corresponding average reduced $\chi^2$ 
value would be 2.3. Shifting down the blue band would not increase its fit to the observed data points. 
Although the model with a Kroupa IMF would then have a reduced $\chi^2$ value of 2.2, the 100\% binaries and
Salpeter IMF models would respectively have reduced $\chi^2$ values of 2.5 and 2.9.

\section{Conclusions} \label{sec:6}

By combining the Munich-Groningen semi-analytic galaxy formation with the SeBa binary population synthesis code
to study the stellar halo WD population, we tried to identify observational features in the halo WD population 
that arise due to their origin in distinct building blocks of the stellar spheroid. 

In the mass versus luminosity diagram of single halo WDs with $G<20$ one main curve for the majority of halo WDs
can be seen and some WDs that are offset from this main curve (see the top panels of Figure~\ref{fig:12}).
The WDs on this main curve all have approximately the same age,
thus if one assumes a main sequence evolutionary lifetime of these WDs, the age of the stellar halo can be derived from 
the WD mass corresponding to this curve. A similar age-determination of the inner halo was suggested by \citet{Kalirai:2012}.
We found that single halo WDs originating in a building block with a significant fraction of young halo stars ($\sim 4$ Gyr old)
in the Solar neighborhood (e.g. B4 from Figure~\ref{fig:11}) will have positions offset from the main curve in this diagram. 
Unfortunately however, WDs that are the result of a binary WD merger in any building block can have the same offset from the main curve.
Thus it will not be possible to assign the offset WDs to a building block of the galactic halo that contains a
larger fraction of young halo stars.
An offset to the other side of the curve is expected for WDs from building blocks with more metal-rich stars 
(see Appendix~\ref{Ap:C}) and again from binary WD mergers, although the former are not expected 
to contribute a significant number of bright WDs to the stellar halo in the Solar neighborhood.

The predicted diagrams of the unresolved binary WD period versus the mass of the brightest WD in these systems 
(the bottom panels of Figure~\ref{fig:12}) are very much alike for the five simulated stellar haloes. 
Therefore, we concluded that the differences between unresolved binary WD populations
originating from ZAMS stars in different bins of the age-metallicity map are no longer
visible in a realistic population of halo WDs. However, there are significant uncertainties in the binary evolution 
at low metallicity that can only be resolved once a larger set of binary WDs at low metallicity has been observed.

The five Aquarius stellar halo WDLFs that we simulated from the combined WD populations in the four
selected building blocks of each stellar spheroid do not differ much from 
each other, mainly because we defined the stellar mass in unevolved stars in our simulation box to equal the 
expected value from the observed mass density by \citet{Fuchs:1998}.
Futhermore, all models assume the same IMF and WD cooling models. 
It is however interesting to compare the WDLF band spanned by these five models with the observed halo WDLF
by RH11 and with the best-fit models of Paper~I. We saw that models with a Kroupa or Salpeter IMF fit the WDLF better
than those with a Chabrier IMF, since the Chabrier IMF can be considered more top-heavy than the Kroupa and Salpeter IMFs
after fixing the halo WD mass in unevolved stars (see Figure~\ref{fig:A2}) which leads to an over-estimation of the 
number of WDs in total in our simulation box. Overall, there is however still quite a good match to the observed WDLF,
especially regarding the fact that we normalized the WDLF independently, i.e. we did not fix our theoretical curve to the
observed one. Furthermore, if we would have taken the lower limit of $\rho_0$, the Aquarius stellar halo WDLFs would fit the observed WDLF 
just as well as the simpler models when these are normalized using our standard value of $\rho_0$ that is 1.5 times larger.

In paper~I it was found that Gaia is expected to detect $\sim$1500 halo WDs. Using cosmologically motivated models of 
the stellar halo of the Milky Way in the Solar neighborhood, we now found $\sim$2200 halo WDs with $G<20$ in our simulation box.
Although this new estimate might be too large, since the number of WDs in some bins of the WDLF is much larger than  
in the observed one by RH11, the total number of known halo WDs will be greatly improved with respect to previous catalogues 
by observations of the Gaia satellite, which will also greatly improve the constraints on the halo WDLF.

\begin{acknowledgements} 
The authors are indebted to the Virgo Consortium, which was responsible for designing and running the halo simulations of the Aquarius Project 
and the L-Galaxies team for the development and maintenance of the semi-analytical code. In particular, we are grateful to Gabriella De Lucia and 
Yang-Shyang Li for the numerous contributions in the development of the code.
PvO thanks the Netherlands Research School for Astronomy (NOVA) for financial support.
ES gratefully acknowledges funding by the Emmy Noether program from the Deutsche Forschungsgemeinschaft (DFG).
AH acknowledges financial support from a VICI grant.
\end{acknowledgements}

\bibliographystyle{aa}
\bibliography{wdpaper}

\begin{thebibliography}{66}
\expandafter\ifx\csname natexlab\endcsname\relax\def\natexlab#1{#1}\fi

\bibitem[{{Abt}(1983)}]{Abt:1983}
{Abt}, H.~A. 1983, \araa, 21, 343

\bibitem[{{Adams} \& {Laughlin}(1996)}]{Adams:1996}
{Adams}, F.~C. \& {Laughlin}, G. 1996, \apj, 468, 586

\bibitem[{{Althaus} {et~al.}(2015){Althaus}, {Camisassa}, {Miller Bertolami},
  {C{\'o}rsico}, \& {Garc{\'{\i}}a-Berro}}]{Althaus:2015}
{Althaus}, L.~G., {Camisassa}, M.~E., {Miller Bertolami}, M.~M., {C{\'o}rsico},
  A.~H., \& {Garc{\'{\i}}a-Berro}, E. 2015, \aap, 576, A9

\bibitem[{{Althaus} {et~al.}(2007){Althaus}, {Garc{\'{\i}}a-Berro}, {Isern},
  {C{\'o}rsico}, \& {Rohrmann}}]{Althaus:2007}
{Althaus}, L.~G., {Garc{\'{\i}}a-Berro}, E., {Isern}, J., {C{\'o}rsico}, A.~H.,
  \& {Rohrmann}, R.~D. 2007, \aap, 465, 249

\bibitem[{{Althaus} {et~al.}(2013){Althaus}, {Miller Bertolami}, \&
  {C{\'o}rsico}}]{Althaus:2013}
{Althaus}, L.~G., {Miller Bertolami}, M.~M., \& {C{\'o}rsico}, A.~H. 2013,
  \aap, 557, A19

\bibitem[{{Althaus} {et~al.}(2009){Althaus}, {Panei}, {Romero}, {Rohrmann},
  {C{\'o}rsico}, {Garc{\'{\i}}a-Berro}, \& {Miller Bertolami}}]{Althaus:2009}
{Althaus}, L.~G., {Panei}, J.~A., {Romero}, A.~D., {et~al.} 2009, \aap, 502,
  207

\bibitem[{{An} {et~al.}(2013){An}, {Beers}, {Johnson}, {Pinsonneault}, {Lee},
  {Bovy}, {Ivezi{\'c}}, {Carollo}, \& {Newby}}]{An:2013}
{An}, D., {Beers}, T.~C., {Johnson}, J.~A., {et~al.} 2013, \apj, 763, 65

\bibitem[{{Annis} {et~al.}(2014){Annis}, {Soares-Santos}, {Strauss}, {Becker},
  {Dodelson}, {Fan}, {Gunn}, {Hao}, {Ivezi{\'c}}, {Jester}, {Jiang},
  {Johnston}, {Kubo}, {Lampeitl}, {Lin}, {Lupton}, {Miknaitis}, {Seo}, {Simet},
  \& {Yanny}}]{Annis:2014}
{Annis}, J., {Soares-Santos}, M., {Strauss}, M.~A., {et~al.} 2014, \apj, 794,
  120

\bibitem[{{Arenou} {et~al.}(2005){Arenou}, {Babusiaux}, {Ch{\'e}reau}, \&
  {Mignot}}]{Arenou:2005}
{Arenou}, F., {Babusiaux}, C., {Ch{\'e}reau}, F., \& {Mignot}, S. 2005, in ESA
  Special Publication, Vol. 576, The Three-Dimensional Universe with Gaia, ed.
  C.~{Turon}, K.~S. {O'Flaherty}, \& M.~A.~C. {Perryman}, 335

\bibitem[{{Chabrier}(1999)}]{Chabrier:1999}
{Chabrier}, G. 1999, \apjl, 513, L103

\bibitem[{{Chabrier}(2003)}]{Chabrier:2003}
{Chabrier}, G. 2003, \pasp, 115, 763

\bibitem[{{Chabrier} {et~al.}(1996){Chabrier}, {Segretain}, \&
  {M'era}}]{Chabrier:1996}
{Chabrier}, G., {Segretain}, L., \& {M'era}, D. 1996, \apjl, 468, L21

\bibitem[{{Cojocaru} {et~al.}(2015){Cojocaru}, {Torres}, {Althaus}, {Isern}, \&
  {Garc{\'{\i}}a-Berro}}]{Cojocaru:2015}
{Cojocaru}, R., {Torres}, S., {Althaus}, L.~G., {Isern}, J., \&
  {Garc{\'{\i}}a-Berro}, E. 2015, \aap, 581, A108

\bibitem[{{Cooper} {et~al.}(2010){Cooper}, {Cole}, {Frenk}, {White}, {Helly},
  {Benson}, {De Lucia}, {Helmi}, {Jenkins}, {Navarro}, {Springel}, \&
  {Wang}}]{Cooper:2010}
{Cooper}, A.~P., {Cole}, S., {Frenk}, C.~S., {et~al.} 2010, \mnras, 406, 744

\bibitem[{{Couchman} \& {Rees}(1986)}]{Couchman:1986}
{Couchman}, H.~M.~P. \& {Rees}, M.~J. 1986, \mnras, 221, 53

\bibitem[{{Croton} {et~al.}(2006){Croton}, {Springel}, {White}, {De Lucia},
  {Frenk}, {Gao}, {Jenkins}, {Kauffmann}, {Navarro}, \&
  {Yoshida}}]{Croton:2006}
{Croton}, D.~J., {Springel}, V., {White}, S.~D.~M., {et~al.} 2006, \mnras, 365,
  11

\bibitem[{{de Bruijne} {et~al.}(2015){de Bruijne}, {Allen}, {Azaz},
  {Krone-Martins}, {Prod'homme}, \& {Hestroffer}}]{de-Bruijne:2015}
{de Bruijne}, J.~H.~J., {Allen}, M., {Azaz}, S., {et~al.} 2015, \aap, 576, A74

\bibitem[{{De Lucia} \& {Blaizot}(2007)}]{De-Lucia:2007}
{De Lucia}, G. \& {Blaizot}, J. 2007, \mnras, 375, 2

\bibitem[{{De Lucia} \& {Helmi}(2008)}]{De-Lucia:2008}
{De Lucia}, G. \& {Helmi}, A. 2008, \mnras, 391, 14

\bibitem[{{De Lucia} {et~al.}(2004){De Lucia}, {Kauffmann}, \&
  {White}}]{De-Lucia:2004}
{De Lucia}, G., {Kauffmann}, G., \& {White}, S.~D.~M. 2004, \mnras, 349, 1101

\bibitem[{{De Lucia} {et~al.}(2014){De Lucia}, {Tornatore}, {Frenk}, {Helmi},
  {Navarro}, \& {White}}]{De-Lucia:2014}
{De Lucia}, G., {Tornatore}, L., {Frenk}, C.~S., {et~al.} 2014, \mnras, 445,
  970

\bibitem[{{Duquennoy} \& {Mayor}(1991)}]{Duquennoy:1991}
{Duquennoy}, A. \& {Mayor}, M. 1991, \aap, 248, 485

\bibitem[{{Font} {et~al.}(2006){Font}, {Johnston}, {Bullock}, \&
  {Robertson}}]{Font:2006}
{Font}, A.~S., {Johnston}, K.~V., {Bullock}, J.~S., \& {Robertson}, B.~E. 2006,
  \apj, 638, 585

\bibitem[{{Fuchs} \& {Jahrei{\ss}}(1998)}]{Fuchs:1998}
{Fuchs}, B. \& {Jahrei{\ss}}, H. 1998, \aap, 329, 81

\bibitem[{{G{\'o}mez} {et~al.}(2013){G{\'o}mez}, {Helmi}, {Cooper}, {Frenk},
  {Navarro}, \& {White}}]{Gomez:2013}
{G{\'o}mez}, F.~A., {Helmi}, A., {Cooper}, A.~P., {et~al.} 2013, \mnras, 436,
  3602

\bibitem[{{Gonzalez} {et~al.}(2015){Gonzalez}, {Zoccali}, {Vasquez}, {Hill},
  {Rejkuba}, {Valenti}, {Rojas-Arriagada}, {Renzini}, {Babusiaux}, {Minniti},
  \& {Brown}}]{Gonzalez:2015}
{Gonzalez}, O.~A., {Zoccali}, M., {Vasquez}, S., {et~al.} 2015, \aap, 584, A46

\bibitem[{{Hawkins} {et~al.}(2015){Hawkins}, {Jofr{\'e}}, {Masseron}, \&
  {Gilmore}}]{Hawkins:2015}
{Hawkins}, K., {Jofr{\'e}}, P., {Masseron}, T., \& {Gilmore}, G. 2015, \mnras,
  453, 758

\bibitem[{{Heggie}(1975)}]{Heggie:1975}
{Heggie}, D.~C. 1975, \mnras, 173, 729

\bibitem[{{Helmi} {et~al.}(2002){Helmi}, {White}, \& {Springel}}]{Helmi:2002}
{Helmi}, A., {White}, S.~D., \& {Springel}, V. 2002, \prd, 66, 063502

\bibitem[{{Helmi} {et~al.}(2003){Helmi}, {White}, \& {Springel}}]{Helmi:2003}
{Helmi}, A., {White}, S.~D.~M., \& {Springel}, V. 2003, \mnras, 339, 834

\bibitem[{{Howes} {et~al.}(2015){Howes}, {Casey}, {Asplund}, {Keller}, {Yong},
  {Nataf}, {Poleski}, {Lind}, {Kobayashi}, {Owen}, {Ness}, {Bessell}, {da
  Costa}, {Schmidt}, {Tisserand}, {Udalski}, {Szyma{\'n}ski}, {Soszy{\'n}ski},
  {Pietrzy{\'n}ski}, {Ulaczyk}, {Wyrzykowski}, {Pietrukowicz}, {Skowron},
  {Koz{\l}owski}, \& {Mr{\'o}z}}]{Howes:2015}
{Howes}, L.~M., {Casey}, A.~R., {Asplund}, M., {et~al.} 2015, \nat, 527, 484

\bibitem[{{Hurley} {et~al.}(2000){Hurley}, {Pols}, \& {Tout}}]{Hurley:2000}
{Hurley}, J.~R., {Pols}, O.~R., \& {Tout}, C.~A. 2000, \mnras, 315, 543

\bibitem[{{Iben} \& {MacDonald}(1986)}]{Iben:1986}
{Iben}, Jr., I. \& {MacDonald}, J. 1986, \apj, 301, 164

\bibitem[{{Isern} {et~al.}(1998){Isern}, {Garcia-Berro}, {Hernanz},
  {Mochkovitch}, \& {Torres}}]{Isern:1998}
{Isern}, J., {Garcia-Berro}, E., {Hernanz}, M., {Mochkovitch}, R., \& {Torres},
  S. 1998, \apj, 503, 239

\bibitem[{{Kalirai}(2012)}]{Kalirai:2012}
{Kalirai}, J.~S. 2012, \nat, 486, 90

\bibitem[{{Kauffmann} {et~al.}(1999){Kauffmann}, {Colberg}, {Diaferio}, \&
  {White}}]{Kauffmann:1999}
{Kauffmann}, G., {Colberg}, J.~M., {Diaferio}, A., \& {White}, S.~D.~M. 1999,
  \mnras, 303, 188

\bibitem[{{Kleinman} {et~al.}(2013){Kleinman}, {Kepler}, {Koester}, {Pelisoli},
  {Pe{\c c}anha}, {Nitta}, {Costa}, {Krzesinski}, {Dufour}, {Lachapelle},
  {Bergeron}, {Yip}, {Harris}, {Eisenstein}, {Althaus}, \&
  {C{\'o}rsico}}]{Kleinman:2013}
{Kleinman}, S.~J., {Kepler}, S.~O., {Koester}, D., {et~al.} 2013, \apjs, 204, 5

\bibitem[{{Komiya} {et~al.}(2007){Komiya}, {Suda}, {Minaguchi}, {Shigeyama},
  {Aoki}, \& {Fujimoto}}]{Komiya:2007}
{Komiya}, Y., {Suda}, T., {Minaguchi}, H., {et~al.} 2007, \apj, 658, 367

\bibitem[{{Kroupa} {et~al.}(1993){Kroupa}, {Tout}, \& {Gilmore}}]{Kroupa:1993}
{Kroupa}, P., {Tout}, C.~A., \& {Gilmore}, G. 1993, \mnras, 262, 545

\bibitem[{{Li} {et~al.}(2010){Li}, {De Lucia}, \& {Helmi}}]{Li:2010}
{Li}, Y.-S., {De Lucia}, G., \& {Helmi}, A. 2010, \mnras, 401, 2036

\bibitem[{{Li} {et~al.}(2009){Li}, {Helmi}, {De Lucia}, \& {Stoehr}}]{Li:2009}
{Li}, Y.-S., {Helmi}, A., {De Lucia}, G., \& {Stoehr}, F. 2009, \mnras, 397,
  L87

\bibitem[{{Lowing} {et~al.}(2015){Lowing}, {Wang}, {Cooper}, {Kennedy},
  {Helly}, {Cole}, \& {Frenk}}]{Lowing:2015}
{Lowing}, B., {Wang}, W., {Cooper}, A., {et~al.} 2015, \mnras, 446, 2274

\bibitem[{{Mestel}(1952)}]{Mestel:1952}
{Mestel}, L. 1952, \mnras, 112, 583

\bibitem[{{Miller Bertolami} {et~al.}(2013){Miller Bertolami}, {Althaus}, \&
  {Garc{\'{\i}}a-Berro}}]{Miller-Bertolami:2013}
{Miller Bertolami}, M.~M., {Althaus}, L.~G., \& {Garc{\'{\i}}a-Berro}, E. 2013,
  \apjl, 775, L22

\bibitem[{{Nelemans} {et~al.}(2001){Nelemans}, {Yungelson}, {Portegies Zwart},
  \& {Verbunt}}]{Nelemans:2001a}
{Nelemans}, G., {Yungelson}, L.~R., {Portegies Zwart}, S.~F., \& {Verbunt}, F.
  2001, \aap, 365, 491

\bibitem[{{Ness} \& {Freeman}(2016)}]{Ness:2016}
{Ness}, M. \& {Freeman}, K. 2016, \pasa, 33, e022

\bibitem[{{Ness} {et~al.}(2013){Ness}, {Freeman}, {Athanassoula},
  {Wylie-de-Boer}, {Bland-Hawthorn}, {Asplund}, {Lewis}, {Yong}, {Lane}, \&
  {Kiss}}]{Ness:2013}
{Ness}, M., {Freeman}, K., {Athanassoula}, E., {et~al.} 2013, \mnras, 430, 836

\bibitem[{{Portegies Zwart} \& {Verbunt}(1996)}]{Portegies-Zwart:1996}
{Portegies Zwart}, S.~F. \& {Verbunt}, F. 1996, \aap, 309, 179

\bibitem[{{Renedo} {et~al.}(2010){Renedo}, {Althaus}, {Miller Bertolami},
  {Romero}, {C{\'o}rsico}, {Rohrmann}, \& {Garc{\'{\i}}a-Berro}}]{Renedo:2010}
{Renedo}, I., {Althaus}, L.~G., {Miller Bertolami}, M.~M., {et~al.} 2010, \apj,
  717, 183

\bibitem[{{Romero} {et~al.}(2015){Romero}, {Campos}, \& {Kepler}}]{Romero:2015}
{Romero}, A.~D., {Campos}, F., \& {Kepler}, S.~O. 2015, \mnras, 450, 3708

\bibitem[{{Rowell} \& {Hambly}(2011)}]{Rowell:2011}
{Rowell}, N. \& {Hambly}, N.~C. 2011, \mnras, 417, 93

\bibitem[{{Salaris} {et~al.}(2010){Salaris}, {Cassisi}, {Pietrinferni},
  {Kowalski}, \& {Isern}}]{Salaris:2010}
{Salaris}, M., {Cassisi}, S., {Pietrinferni}, A., {Kowalski}, P.~M., \&
  {Isern}, J. 2010, \apj, 716, 1241

\bibitem[{{Salpeter}(1955)}]{Salpeter:1955}
{Salpeter}, E.~E. 1955, \apj, 121, 161

\bibitem[{{Salvadori} {et~al.}(2010){Salvadori}, {Ferrara}, {Schneider},
  {Scannapieco}, \& {Kawata}}]{Salvadori:2010}
{Salvadori}, S., {Ferrara}, A., {Schneider}, R., {Scannapieco}, E., \&
  {Kawata}, D. 2010, \mnras, 401, L5

\bibitem[{{Sandage}(1972)}]{Sandage:1972}
{Sandage}, A. 1972, \apj, 178, 1

\bibitem[{{Springel} {et~al.}(2008){Springel}, {Wang}, {Vogelsberger},
  {Ludlow}, {Jenkins}, {Helmi}, {Navarro}, {Frenk}, \& {White}}]{Springel:2008}
{Springel}, V., {Wang}, J., {Vogelsberger}, M., {et~al.} 2008, \mnras, 391,
  1685

\bibitem[{{Springel} {et~al.}(2001){Springel}, {White}, {Tormen}, \&
  {Kauffmann}}]{Springel:2001}
{Springel}, V., {White}, S.~D.~M., {Tormen}, G., \& {Kauffmann}, G. 2001,
  \mnras, 328, 726

\bibitem[{{Starkenburg} {et~al.}(2013){Starkenburg}, {Helmi}, {De Lucia}, {Li},
  {Navarro}, {Font}, {Frenk}, {Springel}, {Vera-Ciro}, \&
  {White}}]{Starkenburg:2013}
{Starkenburg}, E., {Helmi}, A., {De Lucia}, G., {et~al.} 2013, \mnras, 429, 725

\bibitem[{{Starkenburg} {et~al.}(2016){Starkenburg}, {Oman}, {Navarro},
  {Crain}, {Fattahi}, {Frenk}, {Sawala}, \& {Schaye}}]{Starkenburg:2016}
{Starkenburg}, E., {Oman}, K.~A., {Navarro}, J.~F., {et~al.} 2016, ArXiv
  e-prints

\bibitem[{{Suda} {et~al.}(2013){Suda}, {Komiya}, {Yamada}, {Katsuta}, {Aoki},
  {Gil-Pons}, {Doherty}, {Campbell}, {Wood}, \& {Fujimoto}}]{Suda:2013}
{Suda}, T., {Komiya}, Y., {Yamada}, S., {et~al.} 2013, \mnras, 432, L46

\bibitem[{{Toonen} \& {Nelemans}(2013)}]{Toonen:2013}
{Toonen}, S. \& {Nelemans}, G. 2013, \aap, 557, A87

\bibitem[{{Toonen} {et~al.}(2012){Toonen}, {Nelemans}, \& {Portegies
  Zwart}}]{Toonen:2012}
{Toonen}, S., {Nelemans}, G., \& {Portegies Zwart}, S. 2012, \aap, 546, A70

\bibitem[{{Tumlinson}(2010)}]{Tumlinson:2010}
{Tumlinson}, J. 2010, \apj, 708, 1398

\bibitem[{{van Oirschot} {et~al.}(2014){van Oirschot}, {Nelemans}, {Toonen},
  {Pols}, {Brown}, {Helmi}, \& {Portegies Zwart}}]{van-Oirschot:2014}
{van Oirschot}, P., {Nelemans}, G., {Toonen}, S., {et~al.} 2014, \aap, 569, A42

\bibitem[{{van Oirschot} {et~al.}(2017){van Oirschot}, {Starkenburg}, {Helmi},
  \& {Nelemans}}]{van-Oirschot:2017}
{van Oirschot}, P., {Starkenburg}, E., {Helmi}, A., \& {Nelemans}, G. 2017,
  \mnras, 464, 863

\bibitem[{{Wang} \& {Han}(2012)}]{Wang:2012}
{Wang}, B. \& {Han}, Z. 2012, \nar, 56, 122

\end{thebibliography}

\appendix

\section{The return factor} \label{Ap:A}


In this paper we distinguish between unevolved stars, i.e. stars that do not lose any fraction of their mass $M_\mathrm{unev}$ to the ISM,
and evolved stars, which do lose mass to the ISM, thus for which their final mass $M_\mathrm{f, ev}$ does not equal their
initial mass $M_\mathrm{i, ev}$.
We define $R_\mathrm{ev}$ as the fraction of their initial mass that evolved stars lose to the ISM: $M_\mathrm{f, ev} = (1-R_\mathrm{ev}) M_\mathrm{i, ev}$.
The return factor $R$ is defined as the fraction of the initial mass in all stars that is returned to the ISM, 
and the lock-up fraction $\alpha = 1-R$ represents the mass that is locked up in all stars, i.e. not lost to the ISM:
\begin{equation}
\alpha = \frac{ M_\mathrm{unev} + M_\mathrm{f, ev}}{M_\mathrm{unev} + M_\mathrm{i, ev}} = 1 - \frac{ R_\mathrm{ev} M_\mathrm{i, ev}}{M_\mathrm{unev} + M_\mathrm{i, ev}}.
\end{equation}

\begin{figure}
\centering	
	\resizebox{\hsize}{!}{\includegraphics{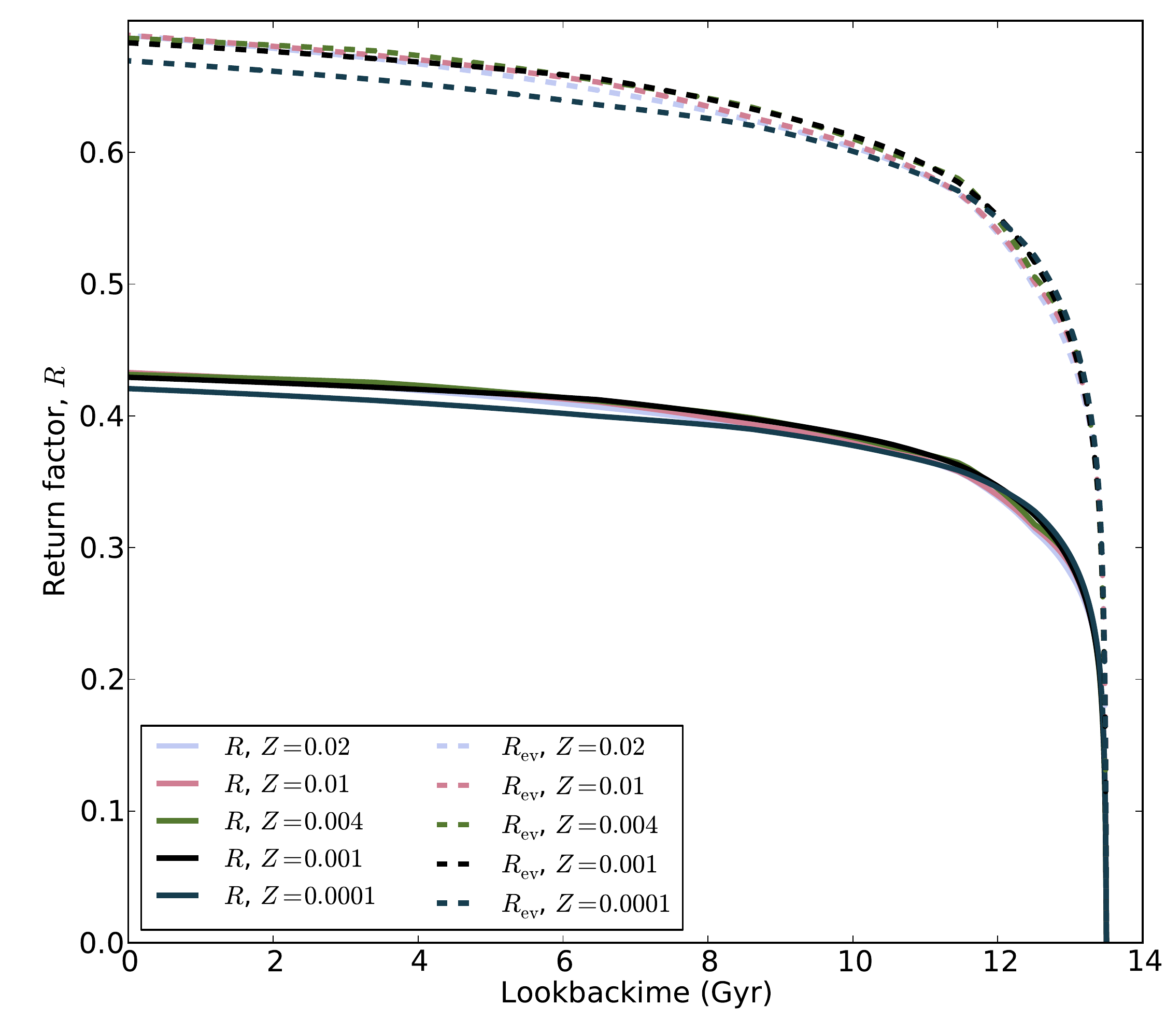}}
		\caption{Return factor of evolved stars (dashed lines) and all stars (solid lines) as a function of
		Lookbacktime, for the five different metallicities used in this study.}
		\label{fig:A1}
\end{figure}

After 13.5 Gyr, only stars above 0.8 $M_\odot$ have evolved, which we define as the boundary mass 
between evolved and unevolved stars. Their mass ratio follows directly from the IMF. 
The \citet{Chabrier:2003} IMF that is used in this paper, is defined as
\begin{equation}
\phi(m) \equiv \frac{\mathrm{d}N}{\mathrm{d}m}\propto 
\left\{
    \begin{array}{ll}
\displaystyle{\frac{1}{m} \exp\left[-\frac{\log_{10}^2 (m/\mu)}{2\sigma^2} \right]} & \ \mathrm{if} \ 0.1 < m \le 1.0 \\ 
&\\
A \ m^{-2.35} & \ \mathrm{if} \ 1.0 \le m < 100. \label{eq:Chabrier}
\end{array} \right.
\end{equation}
with $N$ is the number of stars, $m$ the stellar mass in units of $M_\odot$, $\mu = 0.079$, $\sigma = 0.69$
and the normalization constant
\begin{equation}
A = \exp\left[-\frac{\log_{10}^2 (\mu)}{2\sigma^2} \right] = 0.279.
\end{equation}
For this IMF, the initial mass in evolved stars ($m > 0.8$) is
\begin{eqnarray}
M_\mathrm{i, ev,Chabrier} \propto \int_{0.8}^{1.0} \exp\left[-\frac{\log_{10}^2 (m/\mu)}{2\sigma^2} \right] \mathrm{d}m \nonumber && \\
 + \int_{1.0}^{100} A \ m^{-1.35} \mathrm{d}m && = 0.700 \label{A2} 
\end{eqnarray}
Whereas the mass in unevolved stars ($m\le 0.8$),
\begin{equation}
M_\mathrm{unev,Chabrier} \propto \int_{0.1}^{0.8} \exp\left[-\frac{\log_{10}^2 (m/\mu)}{2\sigma^2} \right] \mathrm{d}m = 0.414. \label{A3} 
\end{equation}
The mass percentage of a single stellar population that is returned to the ISM is of course a function of time, 
that is increasing as the population gets older. We found that its dependance on the binary fraction is 
negligibly small. The effect of the population's metallicity is also small, as we show in Figure~\ref{fig:A1}.
We found that the evolved stars that were born according to a Chabrier IMF lost on average 68\% of their mass,
after evolving them for 13.5 Gyr with the binary population synthesis code SeBa, i.e. $R_\mathrm{ev} = 0.68$,
although the population with $Z=0.0001$ lost 1\% less mass.
This yields
\begin{equation}
R_\mathrm{Chabrier} = \frac{ 0.68\cdot 0.700}{0.414+0.700} = 0.43.
\end{equation}

\begin{figure}
\centering	
	\resizebox{\hsize}{!}{\includegraphics{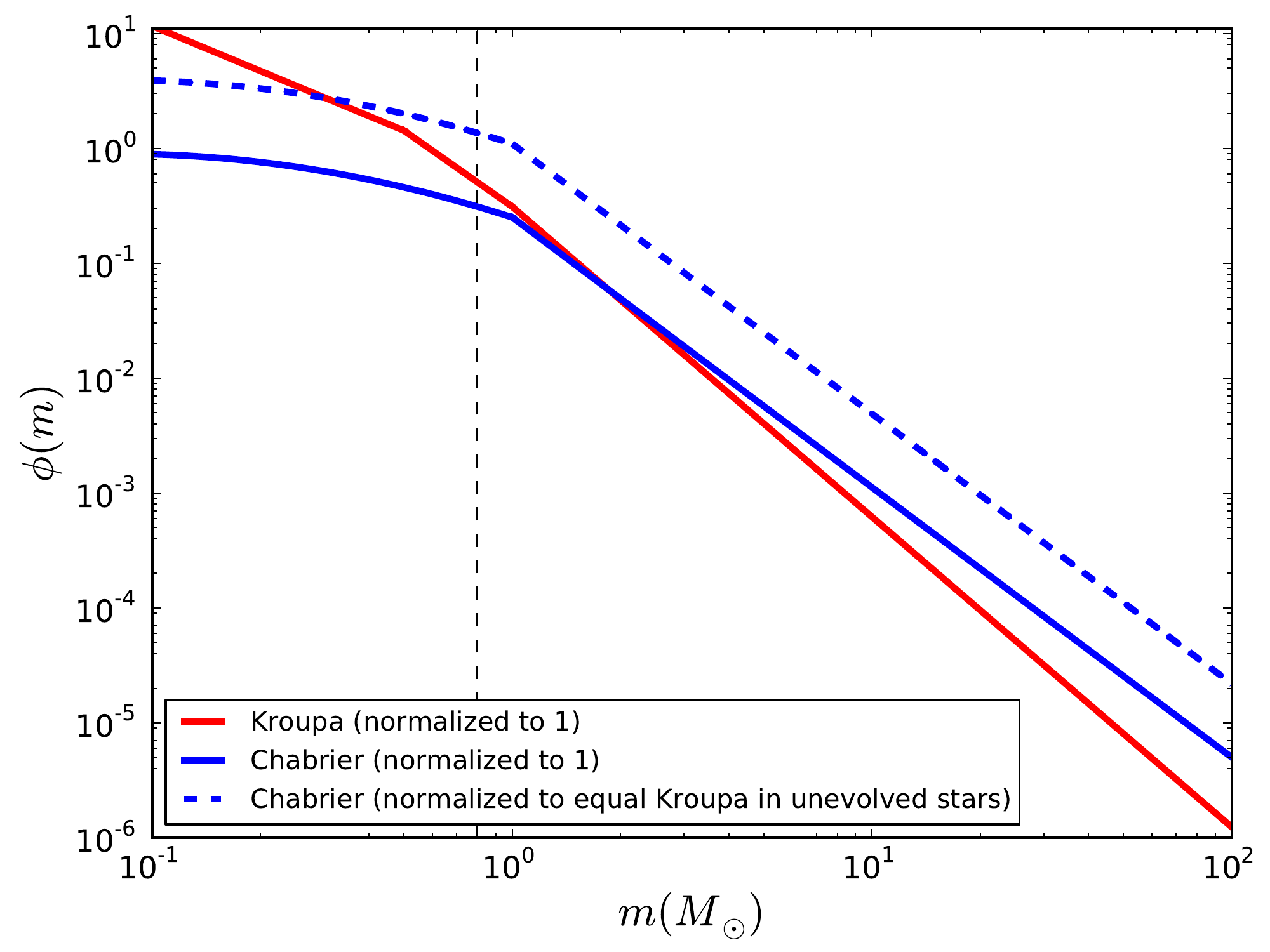}}
		\caption{\citet{Kroupa:1993} versus \citet{Chabrier:2003} Initial Mass Functions,
		normalized such that the integral of the IMF between $m=0.1$ and $m=100$ equals 1
		for each IMF (solid lines). The thick blue dashed line visualises the Chabrier IMF when it is normalized 
		such that the number of unevolved stars equals that number predicted by the Kroupa IMF.
		The thin black dashed line indicates the boundary between evolved and unevolved stars.}
		\label{fig:A2}
\end{figure}

Alternatively, we could have used the \citet{Kroupa:1993} IMF, given by
\begin{equation}
\phi(m) \propto 
\left\{
    \begin{array}{ll}
B \ m^{-1.3} &\qquad \mathrm{if} \ 0.1 \le m < 0.5 \\ 
\quad m^{-2.2} &\qquad \mathrm{if} \ 0.5 \le m < 1.0 \\ 
\quad m^{-2.7} &\qquad \mathrm{if} \ 1.0 \le m < 100 \label{Kroupa}
\end{array} \right.
\end{equation}
with normalization constant 
\begin{equation}
B = \frac{0.5^{-2.2}}{0.5^{-1.3}} = 1.866.
\end{equation}
How this IMF compares to the Chabrier IMF is visualised in Figure~\ref{fig:A2}.
Here, the initial mass in evolved stars
\begin{equation}
M_\mathrm{i, ev,Kroupa} \propto \int_{0.8}^{1.0} \ m^{-1.2} \mathrm{d}m + \int_{1.0}^{100} m^{-1.7} = 1.600
\end{equation}
whereas the mass in unevolved stars 
\begin{equation}
M_\mathrm{i, unev,Kroupa} \propto \int_{0.1}^{0.5} B \ m^{-0.3} \mathrm{d}m + \int_{0.5}^{0.8} m^{-1.2} = 1.624.
\end{equation}
Furthermore, the mass percentage that is returned by evolved stars to the ISM after 13.5 Gyr with a Kroupa IMF is only 62\%, which yields
a much lower return factor,
\begin{equation}
R_\mathrm{Kroupa} = \frac{ 0.62\cdot 1.600}{1.600+1.624} = 0.31.
\end{equation}

\section{Halo WDs in the different bins of the age-metallicity map} \label{Ap:C}

\begin{figure}
\centering	
	\resizebox{\hsize}{!}{\includegraphics{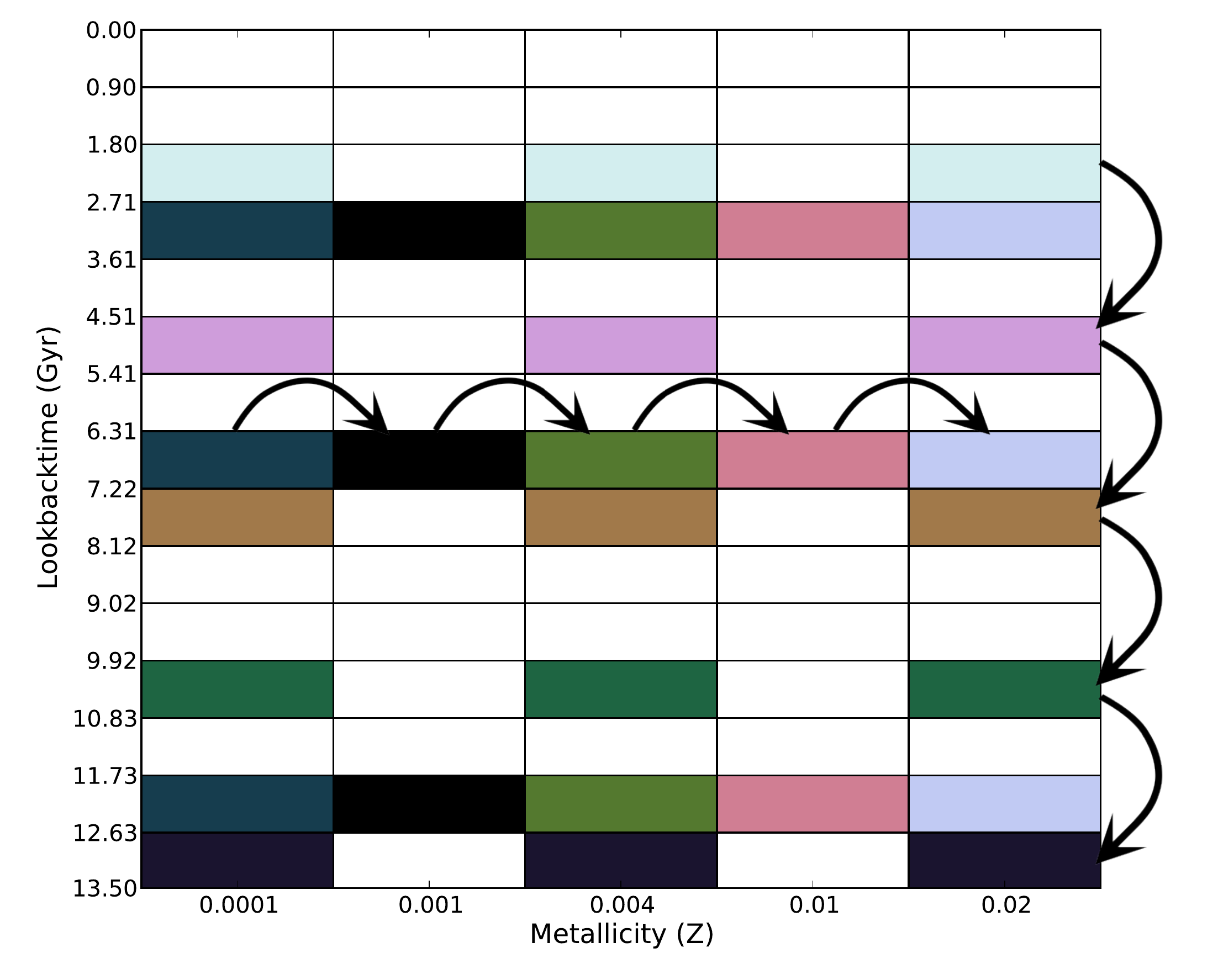}}
		\caption{The $5 \times 15$ bins of the age-matallicity map that is sampled in this study.
		Patch colours match the colours of the points in Figures~\ref{fig:08}, \ref{fig:10} and \ref{fig:14}.
		The vertical arrows indicate the sequence of five colours used in the top panels of 
		these figures, and the horizontal arrows indicate this sequence 
		in their bottom panels. The horizontal colour scheme follows the halo MDF 
		(the green line in Figure~\ref{fig:04}), i.e. the darkest colour is used for the bin where the MDF peaks (Z=0.001).
		The age bins for constant metallicity are also set such that the age bins with increasingly realistic ages for halo 
		stars have darker colours (i.e. darker colours for older stars).		
		To avoid confusion, the age values in the vertical sequence are set to be slightly different from those in 
		the horizontal sequence.}
		\label{fig:07}
\end{figure}

\begin{figure*}
  \centering 
  \resizebox{\hsize}{!}{\includegraphics{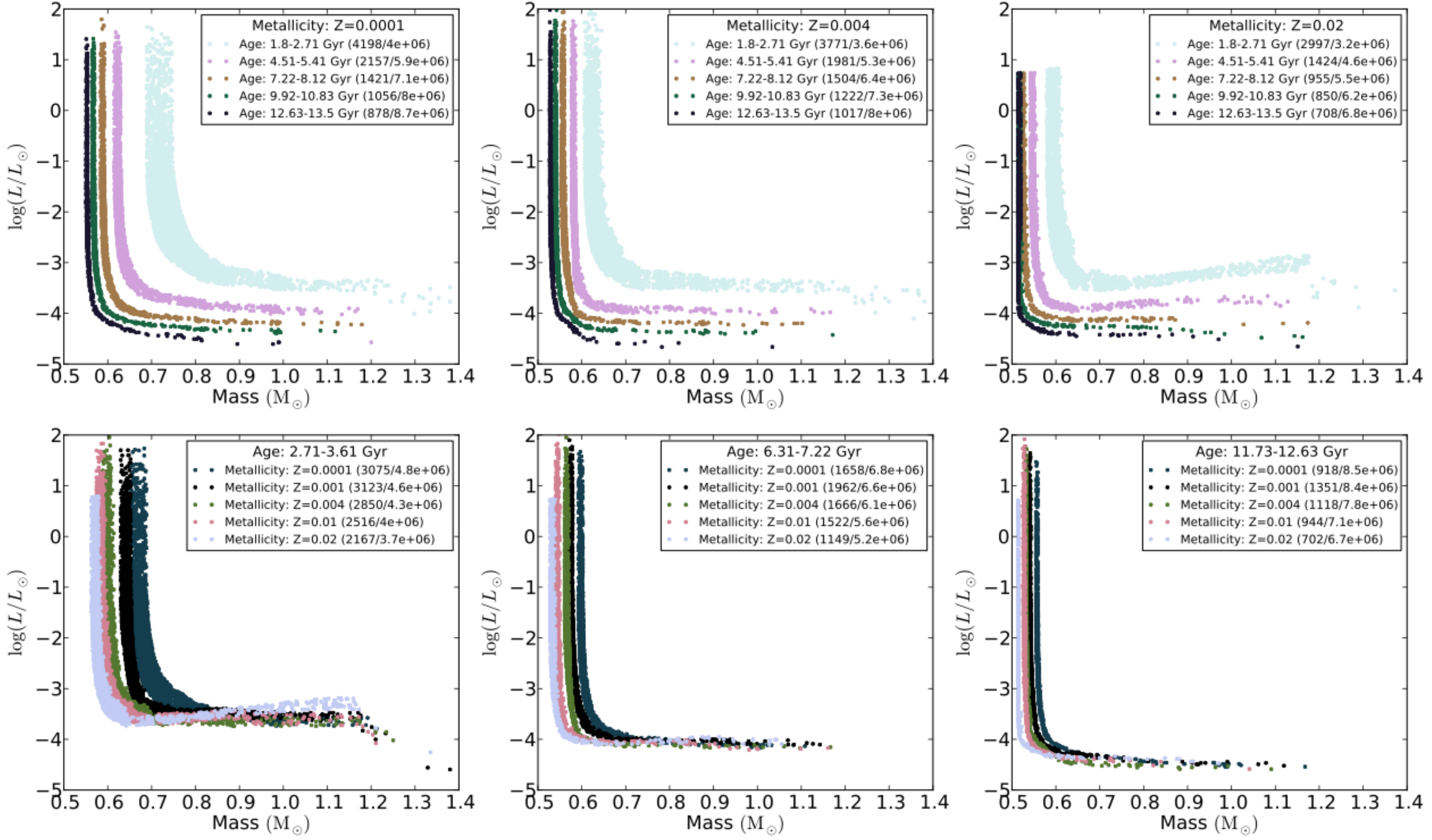}} 
  \caption{Luminosity as a function of stellar mass for single halo WDs in the Solar neighborhood that can be observed with Gaia (G$<$20),
  assuming a single metallicity value for halo stars, for different age ranges (top panels) and a small age spread for halo stars, for different
  metallicity values (bottom panels). }
  \label{fig:08}
\end{figure*}

In this appendix we explore how halo WDs that originate from stars born in different bins of the age-metallicity map 
differ from each other. We make the extreme assumption that $all$ our simulated stars were born in the short timespan of a single age bin 
of the age-metallicity map with a uniform SFR\footnote{The simulated stellar mass in unevolved stars was set to be $1.5\cdot 10^{-4} M_\odot/\mathrm{pc}^3$,
based on the observed value of \citet{Fuchs:1998}. Multiplying this with a factor ($1+0.700/0.414$) to obtain the total mass in ZAMS stars 
(see Appendix~\ref{Ap:A}), and dividing by a timespan of 0.9021 Gyr, we implement a SFR of $ 4.4 \cdot 10^{-13} M_\odot$ yr$^{-1}$ pc$^{-3}$.} 
and that they all have the corresponding metallicity value. 
As can be seen from Figure~\ref{fig:04}, the age-metallicity maps of our stellar haloes and their 
building blocks have $15 \times 5$ bins. Most building blocks span a range of bins, as can be seen in figure~\ref{fig:11}.
The resulting stellar populations therefore do not represent realistic building blocks of the stellar halo,
but they give an idea of the variations between different building blocks due to the different bins of the age-metallicity
map that they span. Figure~\ref{fig:07} shows the bins that we selected to investigate in this section.
The arrows in this Figure indicate sequences of colours that were used in Figures~\ref{fig:08}, \ref{fig:10} and \ref{fig:14}.

Figures~\ref{fig:08}, \ref{fig:10} and \ref{fig:14} all contain six panels. We show WDs with three different metallicities 
in the top three panels of these figures, where five different colours correspond to five different ages. In the bottom three
panels we show WDs with three different ages (taken slightly offset from the ones in the top panels 
to allow for a consistent colouring scheme, see Figure~\ref{fig:07}). Here, five different colours represent five different
metallicities. The colours match those in Figure~\ref{fig:07}. 

In Figure~\ref{fig:08} we show the masses versus luminosity of the single halo WDs with Gaia magnitude $<20$. 
As already remarked in Paper~I, these WDs are expected to follow a narrow curve in this diagram, due to the fact that
most of these brightest WDs have just been formed. Most of them thus have the same mass, which is one to one related
to their initial zero-age main sequence mass and their age, because these are selected not to be in binaries. Compared to
the Gyrs of evolution on the main sequence, the time these WDs need to cool from luminosities above Solar to
$\log(L/L_\odot) < -3$ is a short time (see Figure~\ref{fig:05}). Those WDs that are in these diagrams with lower 
luminosities and higher mass are visible with $G<20$ because they are close to us in terms of distance.

\begin{figure}
  \centering 
  \resizebox{\hsize}{!}{\includegraphics{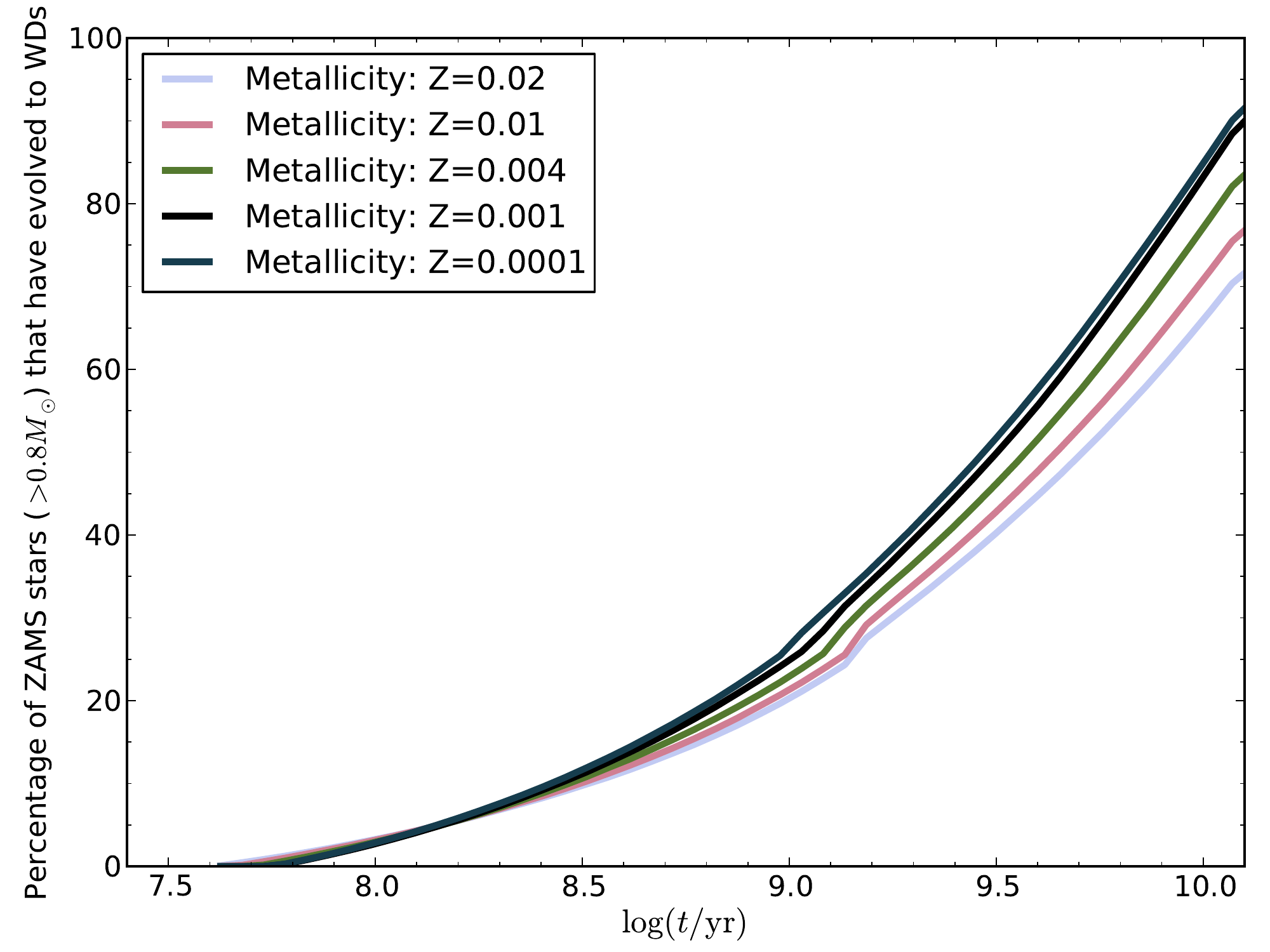}} 
  \caption{Main sequence evolution time scales for the five different metallicities used in this study.
  Colours of the lines are the same as in Figure~\ref{fig:03}. Based on the evolution of $\sim 10^7$ single ZAMS stars 
  with initial mass $> 0.8 M_\odot$, following a Chabrier IMF, for 13.5~Gyr.}
  \label{fig:09}
\end{figure}

It can clearly be seen from the top panels of Figure~\ref{fig:08} that if the halo WD population is younger, 
the WDs with the lowest mass of the population are more massive than those with the lowest mass in an 
older population. The luminosities of the faintest WDs in a young population are furthermore brighter than the
faintest ones in an older population, simply because they had less time to cool. The curves thus shift to the lower left
corner of the panels for increasing population age. The curves also become narrower, because the ratio of the timespan 
of the age-bin ($\sim$0.9 Gyr) over the main-sequence evolution time is larger for younger WDs. Since the 
evolution time of higher mass stars is shorter than that of younger stars, a larger mass range is visible at the
present day if the population is younger. 

The numbers in between brackets in the legend of Figure~\ref{fig:08} indicate the number of single halo WDs with G$<$20 
over the total number of single halo WDs in each selected bin of the age-metallicity map (e.g. including also those with G$\geq$20).
The total number of WDs is obtained by evolving the total number of ZAMS stars 
in our simulation box (see Appendix~\ref{Ap:B}) with SeBa. From these numbers we see that 
there are less WDs with G$\geq$20 in the younger and more metal-rich populations. 

This can be explained by Figure~\ref{fig:09}.
There we plot the percentage of single ZAMS stars with an with initial mass $> 0.8 M_\odot$
that have evolved to WDs (the initial population was assumed to follow a Chabrier IMF), as a function of time ($t$), 
for the five different metallicities used in this study. 
In the younger populations there are less white dwarfs simply because the evolution time of the ZAMS stars was shorter. 
The fact that a more metal-rich population of a certain age (larger than a few 100~Myr, as is the case in Figure~\ref{fig:08}) 
has less white dwarfs in total follows from their slower evolution times, eg. the number of ZAMS stars that have evolved 
to become WDs at that particular age is smaller than for a more metal-poor population.
Although there are less WDs in total in younger populations, the number of bright WDs ($G<20$) is larger than in
older populations of the same metallicity (top panels of Figure~\ref{fig:08}), because the WDs had less time to cool. 

\begin{figure*}
  \centering 
  \resizebox{\hsize}{!}{\includegraphics{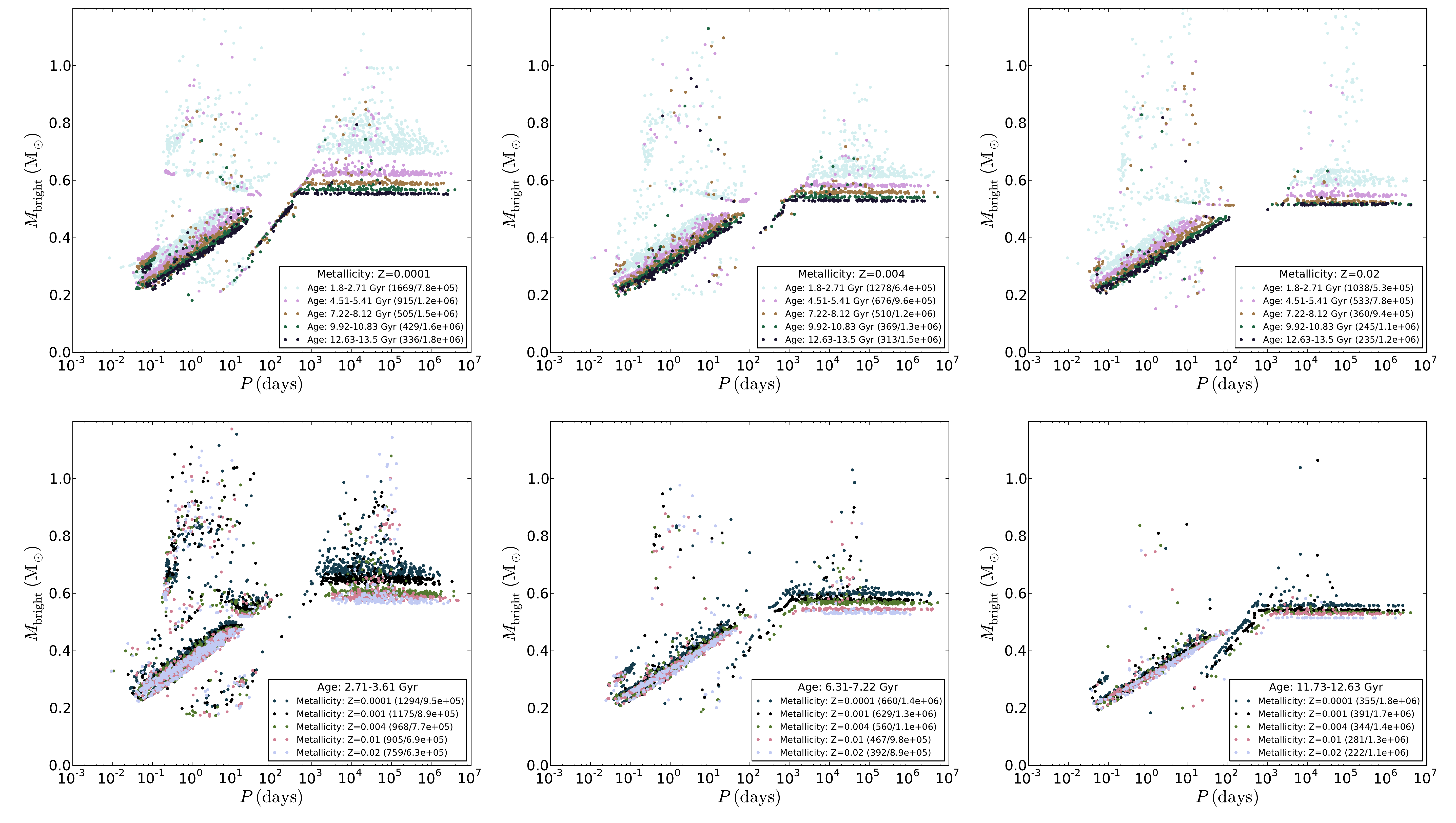}} 
  \caption{Mass of the brightest star in the binary system versus the period of that system in days, for unresolved binary halo WDs in the Solar neighborhood 
  that can be observed with Gaia (G$<$20). Colours for age and metallicity values are the same as in Figures~\ref{fig:07}, \ref{fig:08} and \ref{fig:14}. }
  \label{fig:10}
\end{figure*}

\begin{figure*}
  \centering 
  \resizebox{\hsize}{!}{\includegraphics{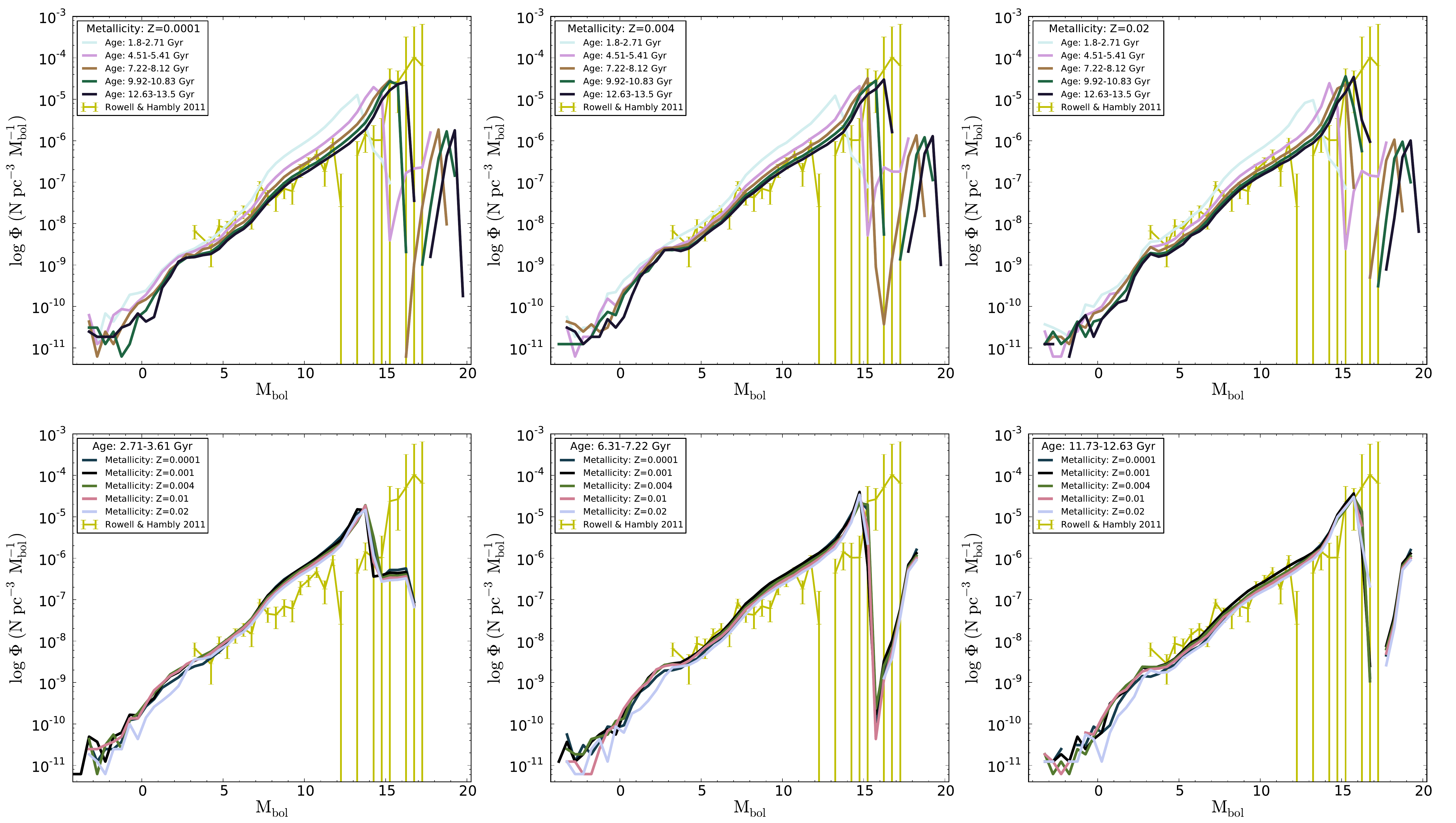}} 
  \caption{Halo WDLFs, based on the assumption that all halo WDs originate from ZAMS stars in a single bin of the age-metallicity map (Figure~\ref{fig:07}).
  Colors for age and metallicity values are the same as in Figures~\ref{fig:07}, \ref{fig:08} and \ref{fig:10}. The yellow lines with errorbars show the observed 
  halo WDLF derived from selected halo WDs in the SuperCOSMOS Sky Survey \citep{Rowell:2011}.
  }
  \label{fig:14}
\end{figure*}

Also in the bottom three panels of Figure~\ref{fig:08} we see that there are less WDs in total in more metal-rich populations 
at a particular age. However, here we see that the number of $G<20$ WDs with Z=0.0001 is lower than that of $G<20$ WDs with Z=0.001.
The difference in the evolution time of the ZAMS stars between these two populations is very small, as can be seen from Figure~\ref{fig:09} and
the total number of single WDs in the simulated populations in the bottom three panels of Figure~\ref{fig:08}. 
It is due to the faster cooling of massive CO WDs with a lower metallicity, as can be seen from the dashed 
lines above cooling times of $10^9$~years in Figure~\ref{fig:05}, that there are less $G<20$ WDs for the the Z=0.0001
population than for the Z=0.001 population in this case.

In Figure~\ref{fig:10} we show the period versus the mass of the brightest star in unresolved binary WDs for the same populations as in Figure~\ref{fig:08}.
As in Paper~I, unresolved binaries are defined as those for which the orbital separation is smaller than 0.3~arcsec,
based on the assumption that two stars in a binary should be separated by at least 0.1$-$0.2 arcsec in order to be spatially resolved by Gaia \citep{Arenou:2005}.
The more recent work of \citet{de-Bruijne:2015} shows that the minimum separation to which Gaia can resolve a close binary probably lies in between 0.23 and 0.70 arcsec,
dependent on the orientation angle under which the binary is observed.
The different aspects of these period versus the mass diagrams were explained for a standard halo model in Paper~I.
Here, we are mainly concerned with variations of this Figure when modelling populations with a different age or metallicity.

In the top panels of Figure~\ref{fig:10} we see that younger populations have systems with $G<20$ in which the brightest WD has a higher mass than
in older populations, similar to the mass trend with population age for single WDs in Figure~\ref{fig:08}. Also the mass range is again larger.
For more metal-rich populations, the period gap (at $M_\mathrm{bright}\sim$0.5~$M_\odot$) shifts towards longer periods. Since single stars of higher metallicity evolve slower (Figure~\ref{fig:09}) 
we capture their binary systems with larger periods because they had less time to evolve towards shorter periods at a particular age.
An interesting feature of Figure~\ref{fig:10} is the (partial) disappearance of the narrow line of systems with $M_\mathrm{bright}\lesssim$0.5~$M_\odot$ moving into the 
above-mentioned period gap for populations with higher metallicity. This also happens in the bottom three panels for the populations that are younger.
The systems on this line have undergone two mass-transfer phases of which the second one was stable, similar to the SNIa progenitors in the single degenerate 
scenario where a non-degenerate companion transfers mass to a WD \citep[for a review, see eg.][]{Wang:2012}. This mechanism does not occur for the most metal-rich
and/or young populations. In the bottom right panel, we see that also the line shifts towards longer periods for more metal-rich populations. 

In between brackets in the legend of Figure~\ref{fig:10}, the number of unresolved binary WDs with G$<$20 is written,
over the total number of unresolved binary WDs in our simulation box, for each of the simulated populations.
In the bottom panels of Figure~\ref{fig:10} we clearly see the effect of population age on the number of unresolved binaries with $G<20$.

Finally, in Figure~\ref{fig:14} we show halo WDLFs for the 30 stellar populations that we investigate in this appendix.
Each panel of Figure~\ref{fig:14} also shows the observed halo WDLF by RH11. 
We applied a correction factor of 0.74 for incompleteness of the observed WDLF to our model lines, 
based on the estimate of RH11\footnote{This correction is a little bit smaller than 
the one that was applied the model lines in Paper~I to compare them with the RH11 data.
There, it was incorrectly assumed that this incompleteness is due to the tangential velocity cut RH11 applied. 
Instead, it should be assigned to their underestimation of the number density of WDs in the Solar neighborhood, 
as they explain in their section~7.4.}.
It is remarkable that the five model lines in the bottom right panel of this Figure fit the data so well, given that we did not
normalize our model lines to the data, as most other authors do. Instead, we normalize the halo WDLF to the corresponding 
observed mass density of local halo (low-mass) main-sequence stars in the Solar neighborhood \citep[see Appendix~\ref{Ap:B}]{Fuchs:1998}. 
From the other panels, it is clear that the effect of age on the WDLF is much larger than that of metallicity. 
Also, we derive from this Figure that the majority of stars in the stellar halo must be at least 9.92~Gyr old in order to match the observed 
data below a reduced $\chi^2$ value of~5. 

\section{Normalization} \label{Ap:B}
From the observed stellar mass density in unevolved halo stars in the Solar neighborhood \citep{Fuchs:1998} we have determined the 
stellar mass corresponding to unevolved stars $M_\mathrm{unev} = 3.6 \cdot 10^7 M_\odot$ in our simulation box
that we use to determine how many stars to simulate (see appendix~A of Paper~I). 
Let $C$ be the normalization constant of the Chabrier IMF, i.e. $M_\mathrm{unev,Chabrier} = 0.414 \ C$ in equation~\ref{A3}.
We have
\begin{equation}
C = \frac{3.6 \cdot 10^7}{0.414} = 8.7 \cdot 10^7 
\end{equation}
and $N_\mathrm{ev} = N_\mathrm{ev,lognormal} + N_\mathrm{ev,Salpeter}$, with 
\begin{equation}
N_\mathrm{ev,lognormal} = C \int_{0.8}^{1.0} \exp\left[-\frac{\log_{10}^2 (m/\mu)}{2\sigma^2} \right] \frac{\mathrm{d}m}{m} = 6.0 \cdot 10^6 \label{B2}
\end{equation}
and
\begin{equation}
N_\mathrm{ev,Salpeter} = C \int_{1.0}^{100} A \ m^{-2.35} \mathrm{d}m = 1.8 \cdot 10^7 \label{B3}
\end{equation}
These numbers are determined from the assumption that all stars are single. We assume that 50\% of the stars are in binaries
however, and that they follow a flat the mass ratio distribution, thus the mass of the secondary is on average half the mass 
of the primary. Therefore, the total number of single stars (which is equal to the total number of binary systems) is equal to 
the sum of the above mentioned numbers (\ref{B2}+\ref{B3}) divided by 2.5.

Alternatively, the semi-analytic model predicts how many stars are born in each bin of the age-metallicity map.
Instead of using the estimate of the mass in unevolved stars in our simulation box from the observed mass density,
we can use the mass (in evolved and unevolved stars) in each bin of the age-metallicity map (initially, i.e.
the present-day mass in each bin divided by $\alpha = 1-0.43$). Dividing this mass by (\ref{A2}+\ref{A3}) 
yields a normalization constant of the IMF for each bin of the age-metallicity map, after which
the same method is used as above to determine the number of evolved stars in our simulation box.

\end{document}